\documentclass[12pt,preprint]{aastex}

\newcommand{\msun}{\ensuremath{\,M_\odot}}
\newcommand{\feh}{\mbox{$\rm{[Fe/H]}$}}
\newcommand{\figref}[1]{\figurename\,\ref{#1}}

\shorttitle{Kepler-424 b}
\shortauthors{Endl et al.}

\begin{document}

\title{Kepler-424 b: A ``Lonely'' Hot Jupiter That Found A Companion.
\footnote{Based on observations obtained with the Hobby-Eberly Telescope, which is a joint project of the 
University of Texas at Austin, the Pennsylvania State University, Stanford University, 
Ludwig-Maximilians-Universit\"at M\"unchen, and Georg-August-Universit\"at G\"ottingen.}
}

\author{Michael Endl}
\affil{McDonald Observatory, The University of Texas at Austin,  
    Austin, TX 78712, USA}
\author{Douglas A. Caldwell}
\affil{NASA Ames Research Center, Moffett Field, CA 94035, USA}
\author{Thomas Barclay}
\affil{Bay Area Environmental Research Inst., 625 2nd St. Ste 209 Petaluma, CA 94952, USA}
\affil{NASA Ames Research Center, Moffett Field, CA 94035, USA}
\author{Daniel Huber}
\affil{NASA Ames Research Center, Moffett Field, CA 94035, USA}
\affil{SETI Institute, 189 Bernardo Avenue, Mountain View, CA 94043, USA}
\author{Howard Isaacson}
\affil{Department of Astronomy, University of California, Berkeley, CA 94720, USA}
\author{Lars A. Buchhave}
\affil{Harvard-Smithsonian Center for Astrophysics, Cambridge, Massachusetts 02138,
USA}
\affil{Centre for Star and Planet Formation, Natural History Museum of Denmark,
University of Copenhagen, DK-1350 Copenhagen, Denmark}
\author{Erik Brugamyer}
\affil{Department of Astronomy, The University of Texas at Austin,  
    Austin, TX 78712, USA}
\author{Paul Robertson}
\affil{Department of Astronomy and Astrophysics / Center for Exoplanets \& Habitable Worlds, Pennsylvania State University, USA}
\author{William D. Cochran, Phillip J. MacQueen}
\affil{McDonald Observatory and Department of Astronomy, The University of Texas at Austin,  
    Austin, TX 78712, USA}
\author{Mathieu Havel}
\affil{NASA Ames Research Center, Moffett Field, CA 94035, USA}
\author{Phillip Lucas}
\affil{Centre for Astrophysics Research, University of Hertfordshire, College Lane, Hatfield AL10 9AB, UK}
\author{Steve B. Howell}
\affil{NASA Ames Research Center, Moffett Field, CA 94035, USA}
\author{Debra Fischer}
\affil{Department of Astronomy, Yale University, USA}
\author{Elisa Quintana}
\affil{NASA Ames Research Center, Moffett Field, CA 94035, USA}
\author{David R. Ciardi}
\affil{NASA Exoplanet Science Institute, California Institute of Technology, Pasadena, CA 91125, USA}

\begin{abstract}

Hot Jupiter systems provide unique observational constraints for migration models in multiple systems and binaries.
We report on the discovery of the Kepler-424 (KOI-214) two-planet system, which consists of a transiting hot Jupiter (Kepler-424b) in a 
3.31-d orbit accompanied by a more
massive outer companion in an eccentric ($e=0.3$) 223-d orbit. The outer giant planet, Kepler-424c, is not detected to transit the host 
star.
The masses of both planets and the orbital parameters for the second planet were determined using precise radial velocity (RV) measurements from the
Hobby-Eberly Telescope (HET) and its High Resolution Spectrograph (HRS). 
In stark contrast to smaller planets, hot Jupiters are predominantly found to be lacking any nearby 
additional planets, the appear to be ``lonely'' (e.g. Steffen et al.~2012). 
This might be a consequence of a highly dynamical past of these systems.
The Kepler-424 planetary system is a system with a hot Jupiter in a multiple system, similar to $\upsilon$~Andromedae. 
We also present our results for Kepler-422 (KOI-22), Kepler-77 (KOI-127; Gandolfi et al.~2013), Kepler-43 (KOI-135; Bonomo et 
al.~2012), and Kepler-423 (KOI-183).
These results are based on spectroscopic data collected with the Nordic Optical Telescope (NOT), the Keck 1 telescope
and HET. For all systems we rule out false positives based on various follow-up observations,  
confirming the planetary nature of these companions. We performed a comparison with planetary evolutionary models which indicate
that these five hot Jupiters have a heavy elements content between 20 and 120~M$_{\oplus}$.
\end{abstract}


\section{Introduction}

The {\it Kepler} mission (Borucki et al.~2010) was designed to determine, for the first time, the
frequency of Earth-sized planets in the classical habitable zone of Sun-like stars. By combining quasi-continuous
photometric monitoring of $\sim 160,000$ stars in the {\it Kepler} search field with the high level of
photometric precision obtained by {\it Kepler}, it also allows us an unprecedented statistical overview 
of the size-distribution of exoplanets. It has been shown by Borucki et al.~(2011) and 
Batalha et al.~(2012) that the vast majority of planets that {\it Kepler} finds are small planets
in the radius range from 1 to 4~R$_{\oplus}$ and that the hot Jupiters, gas giant planets
with radii similar to Jupiter and with orbital periods of a few days, are a very small 
minority of the overall planet population in the Kepler field. Howard et al.~(2012) estimate the
frequency of hot Jupiter candidates in the the {\it Kepler} search volume as 0.5$\pm$0.1\%.
Interestingly, this is only about half the occurrence rate of this type of planets derived from
precise radial velocity (RV) surveys of the solar neighborhood. Wright et al.~(2012) estimate
a local hot Jupiter frequency of 1.2$\pm$0.38\% based on the Lick and Keck RV survey results. 

An interesting aspect of the class of hot Jupiter planets is their apparent lack of additional, nearby planetary 
companions. These Jupiters are ``lonely'' in the sense that searches in the 
{\it Kepler} photometry and using Transit-Timing-Variations (TTVs) do not find any other nearby planets in these
systems (Steffen et al.~2012). This is in stark contrast to smaller and lower mass planets which are
very often found in multi-planet systems (e.g. Latham et al.~2011, Rowe et al.~2014), sometimes in mean-motion resonances.

One hypothesis for the formation of such systems is that the migration of the inner planet is caused by planet-planet scattering (e.g. Rasio \& 
Ford~1996) rather than by disk-driven tidal migration (e.g. Goldreich \& Tremaine~1980). 
The large eccentricity of the inner planet's orbit, pumped up by a close encounter with a 
massive second giant
planet in the system, is subsequently circularized by tidal forces at small separations to the host star and establishes the hot Jupiter orbit.
The outer planet's orbit remains eccentric since tidal interactions are too weak at these separations. However, the planet-planet scattering
process might be quite violent and potentially could lead more often to the loss of the inner planet by falling into the star than by forming a
hot Jupiter. Conversely, the strong planet-planet interactions could also lead to the ejection of the outer planet, leaving behind a single
hot Jupiter planet. Observational footprints of planet-planet interactions include moderate to high eccentricities of the outer planet (if one exists),
an inclined orbital plane of the
inner planet with respect to the stellar spin axis, and significant non-zero mutual inclinations. Interestingly, McArthur et al.~(2010)
measured a mutual inclination of $\sim 30^{\circ}$ of the orbits of $\upsilon$~And c ($P=241$~d) and d ($P=1282$~d). While the inclination of
$\upsilon$~And~b ($P=4.6$~d) remains unknown, this could indicate a very dynamical past of this system.

Moreover, hot Jupiter planets appeared to rarely have outer giant planets at longer 
periods found by RVs (e.g. discussion in Bryan et al.~2012). This paucity of additional planets in hot Jupiter systems was
usually interpreted as evidence for a significant dynamical evolution in the past, that cleared out all or most of the other (detectable) planets.
Systems where a hot Jupiter is found in a multi-planet system (e.g. the $\upsilon$~Andromedae system) are therefore
ideal testbeds to search for evidence for dynamical evolution of the system.

Knutson et al.~(2014) presented first results of their RV survey of hot Jupiter systems to detect additional massive companions that could
dynamically effect the close-in giant planets. They estimate a frequency of $51\pm10$\% for high-mass companions (from giant planets to brown dwarfs)
at larger orbital separations. They also did not find a difference in the occurrence rates of systems with misaligned or eccentric orbits and 
well-aligned and circular orbits.

In this paper we present the data and their analysis performed 
by members of the {\it Kepler} team and follow-up observers 
at McDonald Observatory (HET \& 2.7~m telescope), La Palma (NOT), Mauna Kea (Keck I) and Kitt Peak (WIYN) 
for five Kepler-Object-of-Interest (KOI): KOI-22, KOI-127, KOI-135, KOI-183, and KOI-214.
We highlight the Kepler-424 (= KOI-214) system, a new 2-planet system, where we have found a second (outer) giant planet in this system 
using precise RV measurements.

\section{{\it Kepler} Photometry and Transit Signatures} 

We analyzed 16 quarters (Q1-Q16) of highly precise {\it Kepler} photometry to determine the transit parameters. The long cadence data, with 
30 minute sampling, was processed using the standard {\it Kepler} pipeline (Jenkins et al. 2010). The {\it Kepler} photometry is publicly 
available at the Mikulski Archive for Space Telescopes (MAST\footnote{http://archive.stsci.edu/kepler/data\_search/search.php})
at the Space Telescope Science Institute.

These five targets were flagged early on in the {\it Kepler} mission as Kepler-Object-of-Interest with transit periods 
of 7.89 days (KOI-22), 3.58 days 
(KOI-127), 3.02 days (KOI-135), 2.68 days (KOI-183), and 3.31 days (KOI-214). 
All transit signals were deep ($5,000 - 18,000$ ppm), high signal-to-noise (S/N) 
events typical of transits of giant planets. 

The data validation (DV) reports for each KOI include an analysis of the centroid fit to the pixel response 
function (PRF) and compare the centroids of in-transit to the out-of-transit images. 
A detailed description of the overall method can be found in Bryson et al.~(2013).
A significant 
shift of the centroids in the difference images can indicate a false positive, in particular a false positive caused by unresolved background eclipsing 
binaries. But this is not the only reason for a centroid shift, also crowding of the field can produce a significant offset during transit, as in the case of 
Kepler-15 (KOI-128) that has a measured centroid offset of 0.1 arcsec with a significance of $6.5\sigma$ (see discussion in Endl et al.~2011).   

For all five targets we have DV reports that cover multiple quarters (Q1-Q12). 
Three of them have non-significant mean centroid offsets: KOI-127 (0.16$\sigma$), KOI-135 (0.24$\sigma$) and KOI-214 (0.85$\sigma$). 
KOI-183 has a low significant centroid offset of 2.3$\sigma$ and KOI-22 just reaches the warning threshold of 3$\sigma$ by showing a multi-quarter
offset of 0.0223 arcsec at a significance of 3.09$\sigma$. 
KOI-22 is therefore a possible false positive that required additional high-angular resolution 
imaging for validation (see section\,\ref{images}). We note that KOI-22 also has the highest false positive probability 
(FPP) of 3\% of these five KOIs as estimated in Morton \& Johnson~(2011), with the remaining four all have an FPP of 1\%. 
However, this is still below the 6\% mean FPP of the 1235 KOIs considered in Morton \& Johnson~(2011). Santerne et al.~(2012) estimated a
relatively high general false positive rate of 35\% for hot Jupiter KOIs, which means that in a sample of five we could expect one false
positive. Extra care has to be taken to obtain all required follow-up observations to rule out a false positive scenario.

\section{Follow-up Observations}

\subsection{Reconnaissance Spectroscopy}

One of the first steps in the {\it Kepler} Follow-up Observing Program (FOP) for KOIs is the acquisition of reconnaissance spectra, which 
allow a first spectroscopic determination of T$_{\rm eff}$, log $g$, and $v \sin i$ of the host star. We usually take two spectra at opposite quadratures of 
the expected RV orbit (based on the transit ephemeris) to quickly rule out grazing eclipsing binaries as a source of false positives.

Reconnaissance spectra of the five KOIs were obtained with the Tull C\'oude Spectrograph (Tull et al.~1995) at the Harlan J. Smith 2.7\, m Telescope at McDonald
Observatory and with the FIbre--fed \'Echelle Spectrograph (FIES) at the 2.5\,m Nordic Optical Telescope (NOT)
at La Palma, Spain (Djupvik \& Andersen\,2010) and with the Hamilton Spectrograph at Lick Observatory. For all five KOIs, the result of this initial
reconnaissance showed that these objects are slowly rotating, single, solar-type stars, that are suitable for continued follow-up observations with the goal
to confirm the planet and determine its mass and density.

\subsection{Imaging}
\label{images}

\begin{figure}
\includegraphics[trim=20 150 10 200,clip,scale=0.9]{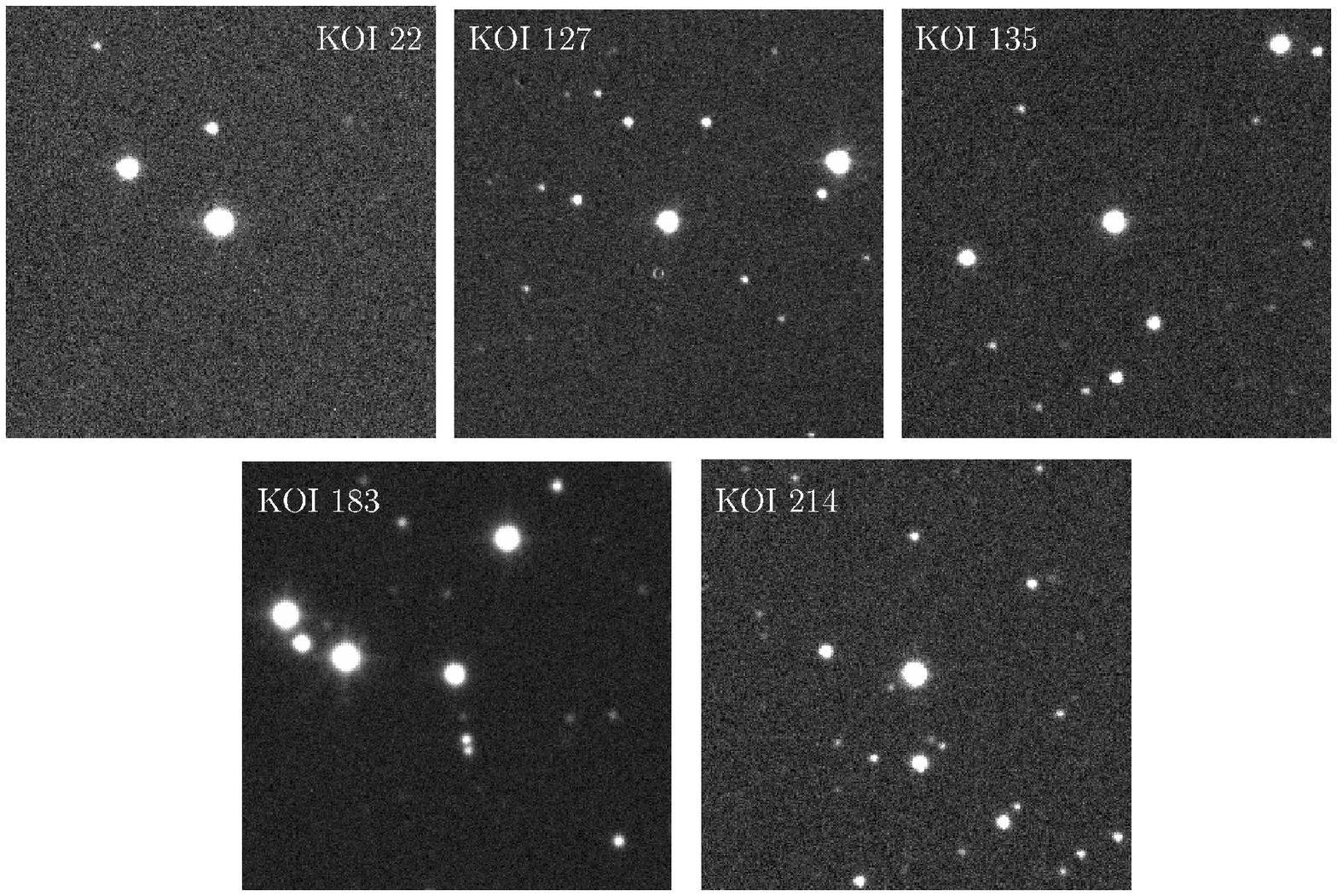}
\caption{
UKIRT $1\times1$ arcminute J band images of the fields around the five systems. The KOI
is at the center of each image. North is up and East is left. None of the KOIs has a nearby
contaminating star.
\label{ukirt}}
\end{figure}

Figure~\ref{ukirt} shows the UKIRT $J$ band images of these five targets taken in July 2009 as part of the UKIRT
survey of the {\it Kepler} field\footnote{http://keplergo.arc.nasa.gov/ToolsUKIRT.shtml}.
The images have full-width-half-maxima of 0.75 to 1.0 arcsec. Sources down to 
$J=19$ mag (Vega system) are detected with near 100\% completeness.  The ring
seen to the south of KOI-127 is a an electronic cross-talk artifact caused by a very
bright star located ~2 arcminutes to the west. Inspection of all images reveal a single star with no
nearby bright contaminating stars.

For KOI-22, the possible false positive candidate from the DV report, we obtained speckle observations at the 
WIYN 3.5\,m telescope on Kitt Peak using 
the same procedures as described in Howell et al.~(2011). No companion (or background) star
was revealed by these $2.76\times2.76$ arcsec images down to a magnitude limit below the target star of 4.0 magnitudes in $R$ and
4.0 magnitudes in $V$. KOI-22 was also observed by Robo-AO (Law et al.~2013) and no new companions were detected within 
2.5 arcsec. We also obtained J-band images for this KOI with the Palomar 200-inch telescope and its adaptive optics system. These images
have a FWHM resolution of 0.1 arcsec and also do not show any additional star within 1 arcsec down to a level of 5 delta magnitudes and
down to 7.5 delta magnitudes at 2 arcsec radial distance. 
Given that the KOI-22 transits are deep (11.29 mmag) and that both AO and speckle did not detect any nearby
stars, we regard this target as a viable planet candidate.

\subsection{Precise Radial Velocity Measurements}

We obtained precise RV follow-up measurements for KOI-214 with the 
Hobby-Eberly-Telescope (HET) and its HRS spectrograph (Tull~1998) at McDonald Observatory.
We employed the same instrumental setup ($R=30,000$) and data 
reduction pipeline as for Kepler-15 (Endl et al.~2011). From July 2010 to April 2013 we observed KOI-214 30 times 
with the iodine cell for the RV determination and once without the cell to acquire a stellar
``template'' spectrum. Exposure times range from 20 to 40 minutes per spectrum. It became
quickly obvious that the expected 3.31-d RV signal is modulated by a second, longer-period signal.  
Owing to continued monitoring with the HET, we were able to characterize the second signal as
due to a more massive, outer planet.

Radial velocity data for KOI-22 were taken with Keck/HIRES with a resolving power of $R=60,000$, using an instrumental setup similar
to the California Planet Search (e.g. Howard et al.~2010). 
The iodine cell setup was utilized to monitor real time instrumental variations relevant to measuring precise radial velocities.  
Exposures taken in 2010 and 2011 utilize a sky subtraction technique that improves RV precision by removing contaminating moonlight from the spectra. 
Exposures taken in 2009 do not have sky subtraction, and are less precise that those taken in spring 2010 onward.
The HIRES data were reduced using the same methods as described in Batalha et al.~(2011).

Spectroscopic observations of KOI-127 and for KOI-135
were obtained using the fiber-fed Echelle Spectrograph (FIES) on the 2.5\,m
Nordic Optical Telescope (NOT). We acquired 10 spectra of
KOI-127 between 2 July and 16 October 2010 and 14 FIES spectra of
KOI-135 between 5 July and 20 October 2010.
We used the medium and the high--resolution fibers ($1\farcs3$ projected
diameter) with resolving powers of $\lambda/\Delta\lambda \approx 46,\!000$
and $67,\!000$, respectively, giving a wavelength coverage of $\sim
3600-7400$\,\AA\@. We used the wavelength range from approximately $\sim
3900-5800$\,\AA\@ to determine the radial velocities. The exposure time was
between 45 and 60 minutes yielding an average S/N per
resolution element of 32 and 36, respectively, near the MgB-triplet. The
procedures used to reduce the FIES spectra and extract the radial velocities
are described in Buchhave et al.~(2010).

We also obtained additional precise RV measurements with HET/HRS for KOI-127 and KOI-183. 
From August 2010 to April 2011 we collected 9 spectra for KOI-127 with exposure times ranging from 20 to 30 minutes.
KOI-183 was observed 16 times from July 2010 to June 2012 with exposure times between 20 and 40 minutes. 

\begin{figure}
\includegraphics[angle=270,scale=0.55]{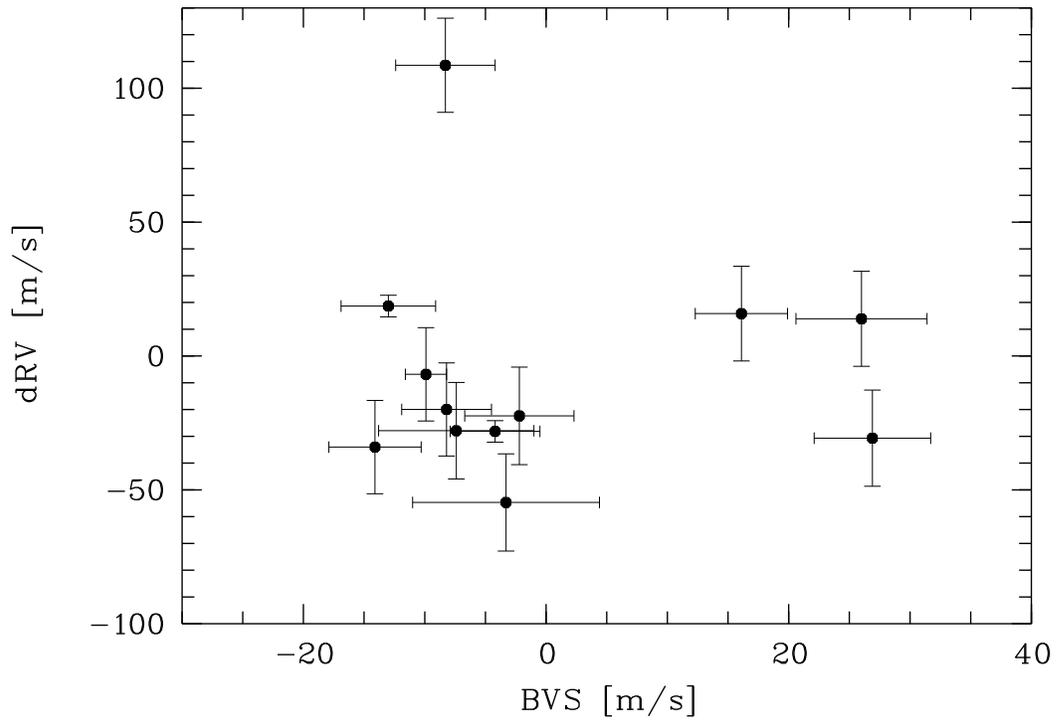}
\caption{Correlation of line bisector velocity span (BVS) measurements and RV results for Kepler-422 (KOI-22), based on the Keck/HIRES 
spectra.
The probability that these two quantities are uncorrelated is 96.5\%. 
\label{bvs}}
\end{figure}

Tables\,\ref{rv214} to \ref{rv183} list all precise RV data for the five KOIs acquired with the three different telescopes and spectrographs. 
These measurements are purely differential, and not absolute RV measurements. 
The data for KOI-22, KOI-127, KOI-135 and KOI-183 all show Keplerian motion consistent in phase with a single transiting giant planet. 
This further strengthens the case that KOI-22, with a centroid offset of 3.09\,$\sigma$, is not a false positive.
Nevertheless, we also performed a line bisector analysis for KOI-22 using the HIRES spectra 
and searched for a correlation of the bisectors with the RV data (Figure\,\ref{bvs}). 
The linear correlation coefficient is -0.014 and the probability of zero-correlation is 96.5\%. 
The total evidence that the deep transit signal seen for KOI-22 is due to a planet includes now (1) a lack of nearby stars detected by AO and speckle 
imaging, (2) the precise RV data in amplitude and phase consistent with a planetary signal, and (3) the lack of gross line bisector variability or
correlation with the photometric phase. We therefore 
conclude that KOI-22.01 is confirmed as a transiting hot Jupiter, Kepler-422b, and KOI-183.01 is confirmed as Kepler-423 b. 
KOI-127.01 and KOI-135.01 were previously confirmed as Kepler-77b and Kepler-43b by Gandolfi et al.~(2013) and Bonomo et al.~(2012).
 
\begin{figure}
\includegraphics[angle=270,scale=0.6]{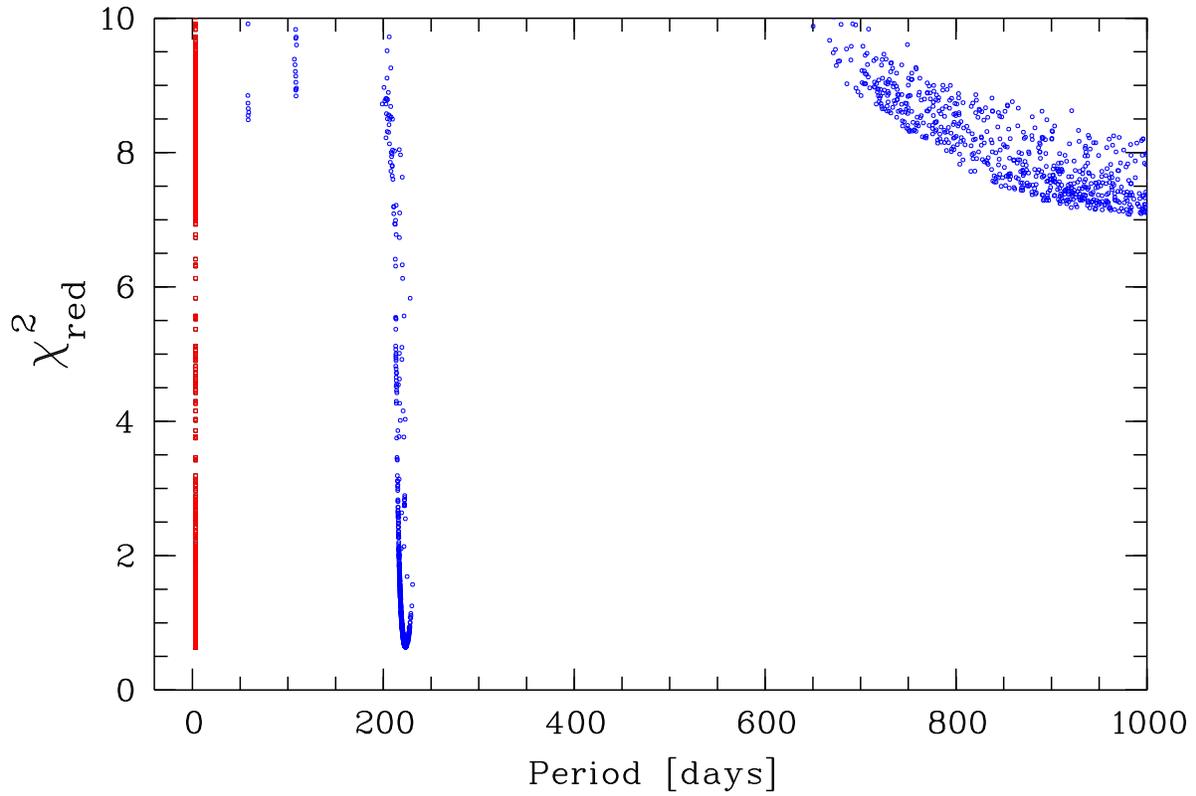}
\caption{Genetic algorithm results for the Kepler-424 RV data using a 2-planet model. The $\chi^{2}_{\rm red}$ values
of 80,000 different 2-planet models are
shown as a function of orbital period. There is a clear minimum at a period of $P \sim 220$~days for the second planet. Period values for
the second planet are displayed as blue circles, while the hot Jupiter is shown as small, red boxes (fixed at the 3.31-d transit period).
\label{gen}}
\end{figure}

For the KOI-214 RV data we used a genetic algorithm to explore the parameter space for a 2-planet system. We fixed the period of the 
inner planet to the transit period \& phase and assumed zero eccentricity for the hot Jupiter, while letting all orbital parameters vary 
freely for the
second companion. We performed 80,000 iterations of the
genetic algorithm. In each iteration, the starting parameters of the model are randomly selected (except the parameters that are fixed for planet b)
and then allowed to ``evolve'' until a local $\chi^{2}$ minimum is reached. By using a large number of iterations we can map the
$\chi^{2}$ surface for any given system without being trapped in local minima.
Figure\,\ref{gen} displays the results in the $\chi^{2}$-period plane for all solutions with $\chi^{2}_{\rm red} \leq 10$ and
$P_{\rm planet\,c} \leq 1000$~days. We find a clear global $\chi^{2}$ minimum around a period of 220 days for the second planet.
All solutions close to this $\chi^{2}$ minimum have a moderate
eccentricity of $0.3-0.35$. We confirm these results later with
a simultaneous Markov Chain Monte Carlo (MCMC) analysis of the KOI-214 photometry and RV measurements (see section\,\ref{mcmc}). 
KOI-214.01 is now confirmed as Kepler-424b. 
A search in the {\it Kepler}
photometry for transits of the second planet, Kepler-424c, was unsuccessful.

\subsection{Host Star Characterization}

Precise stellar parameters for the five KOI host stars were derived using either SME (Valenti \& Piskunov~1996), SPC (Buchhave et al.~2012) or MOOG (Sneden 
1973) on high S/N spectra that do not contain any iodine lines (in the case of HRS and HIRES).
Table\,\ref{stars} summarizes the parameters we determined for the five KOI host stars. For KOI-214 we used all three tools on a Keck/HIRES and a 
HET/HRS spectrum. The reported values for KOI-214 represent the mean and standard deviation from these 3 different approaches.
All five KOIs are solar-type stars with near solar temperatures and surface gravities, and super-solar metallicities ranging from [Fe/H]=+0.23 to +0.44 dex, as 
expected from giant-planet/metallicity relationships from Fischer \& Valenti~(2005) and Santos et al.~(2005).
 
\section{Modeling of Photometry and RV data}
\label{mcmc}
We modeled the light curve and RV measurements simultaneously with a full Keplerian model (Rowe et al. 2014). The transit shapes were described by a an 
analytic 
limb darkened transit using a quadratic limb darkening law. The parameters used in the fit were the mean stellar density, two limb darkening parameters, the orbital period of the 
planet, the mid-point time of the transit, the impact parameter of the planet at mid-transit, the planet-to-star radius ratio, two parameters controlling the 
eccentricity, $e \sin \omega$ and $e \cos \omega$ where $e$ is the eccentricity and $\omega$ is the argument of periastron, and the radial velocity semi-amplitude ($K$). 
Two additional noise parameters (one for the photometry and one ``jitter'' parameter for the RVs) were included that were added in quadrature with the formal 
uncertainty to account for the limitation in the model such as 
additional stellar variability. Finally, we also included two nuisance parameters to account for photometric and RV offset from zero (the RV parameter is often known as $\gamma$). 
In the case of Kepler-424, for the non-transiting planet we set the planet radius to zero and the impact parameter to zero which enabled 
us to simultaneous model both planets 
in the RV model but have only the transiting planet impact the light curve model.

The likelihood of the data was calculated assuming a Gaussian function and we included priors on the limb darkening and stellar density based on the spectroscopically 
derived stellar parameters. We additionally included a prior on the eccentricity of $e^{-1}$ which accounts for a bias induced by 
parameterizing in terms of the eccentricity 
vectors (Eastman et al. 2013). 
The probability of each realization of the model is the product of the likelihood and the priors.

We integrated the probability space using the emcee (Foreman-Mackey et al. 2013) implementation of an affine invariant sampler  
(Goodman \& Weare 2010) which we ran with 500 chains and 20,000 steps of each chain (the procedure is similar to that described in Barclay et al.~2013). 
The chains were all well mixed and converged upon a consistent distribution in each model parameter.

Host star radii and masses were determined by fitting the spectroscopic
temperatures and metallicities (Table\,\ref{stars}) combined with the stellar density derived from the
transit model to stellar interior models. We used two sets of models: evolutionary
tracks from the BaSTI database (Pietrinferni et al.~2004) interpolated to a step-size of
0.01 \msun\ in mass and 0.02\,dex in \feh, and isochrones from the Dartmouth database
~(Dotter et al.~2008) interpolated to a step-size of 0.5\,Gyr in age and 0.02\,dex in \feh.
For BaSTI we used canonical models (no convective-core overshooting), and for
both BaSTI and Dartmouth solar-scaled $\alpha$ abundances were adopted.
For both sets of models, the best-fit parameters and uncertainties were derived through
Monte-Carlo simulations as described in Huber et al.~(2013).
The resulting radii and masses agreed within the formal $1\sigma$ uncertainties in
all cases. Since the Dartmouth grid extended to higher metallicities (+0.56\,dex),
we adopted the Dartmouth stellar properties as our final values, and added in
quadrature the difference to the BaSTI solutions to the formal uncertainties of the
Dartmouth solutions.

\begin{figure}
\begin{center}
\includegraphics[angle=0,scale=0.62]{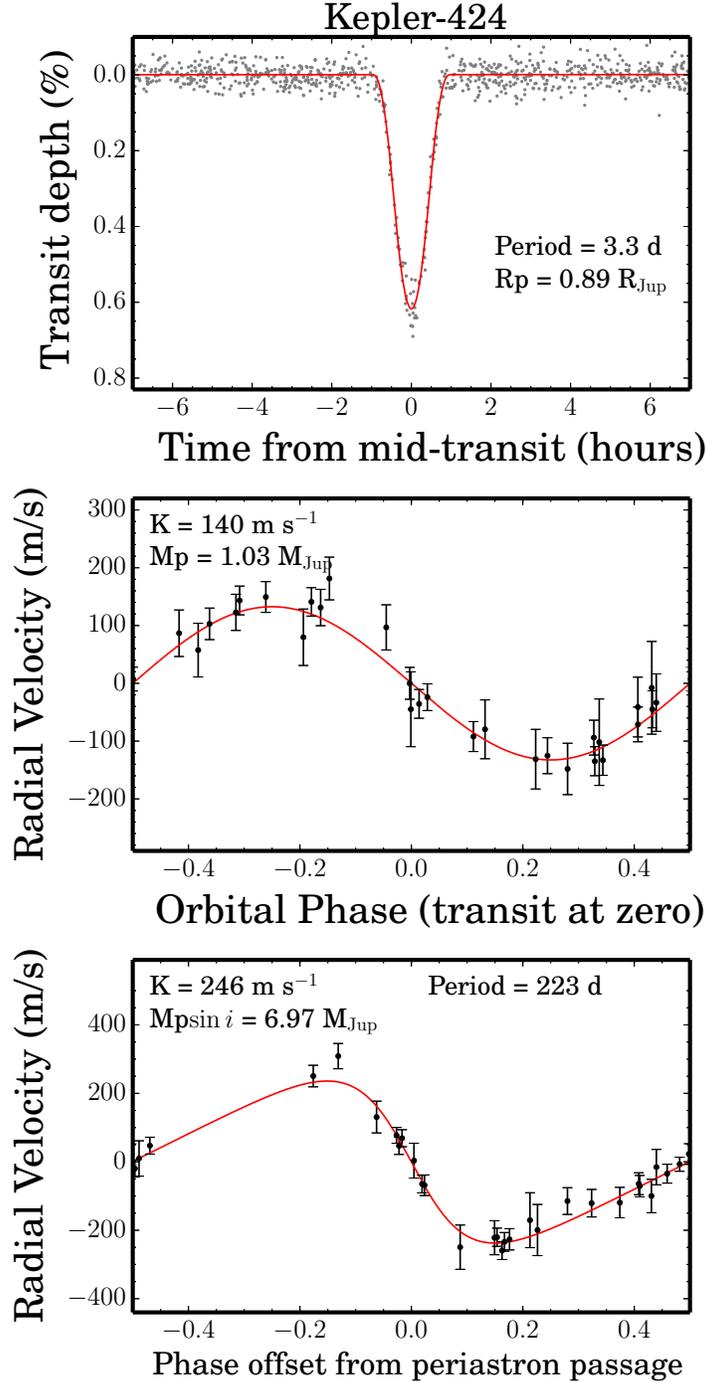}
\end{center}
\caption{Phase-folded {\it Kepler} photometry of Kepler-424 (KOI-214) along with the best-fit transit model from the 
MCMC analysis (top panel). The HET RV data and the best-fit Keplerian orbital solution is shown
for planet b in the middle panel, and for planet c in the bottom panel (in both cases the orbit due to the
other planet is subtracted). The RV uncertainties include an additional RV$_{\rm jitter}$ of 6.6\,m\,s$^{-1}$ from
the highest probability model from the MCMC analysis.
\label{214}}
\end{figure}

\begin{figure}
\includegraphics[angle=0,scale=0.55]{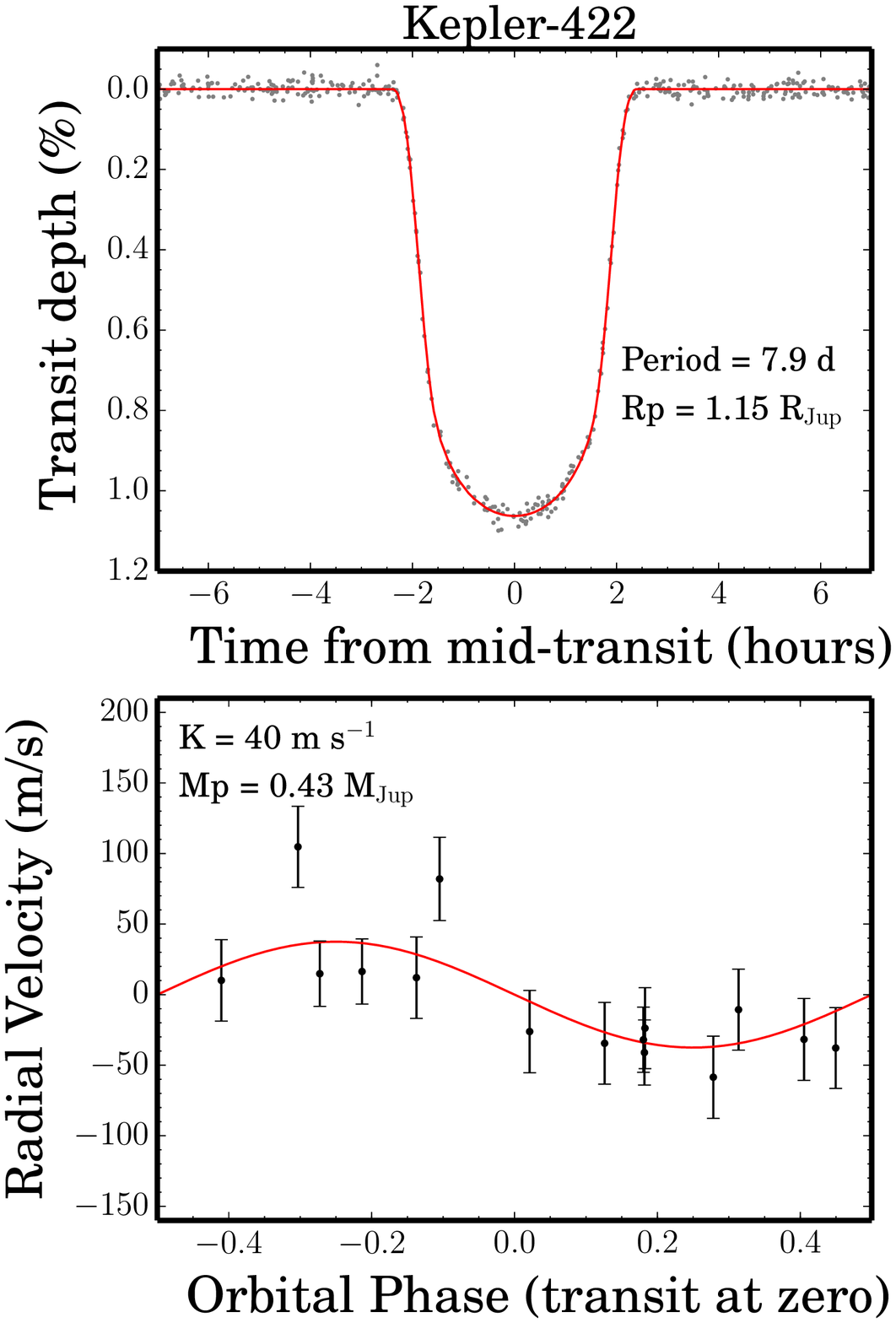}
\includegraphics[angle=0,scale=0.55]{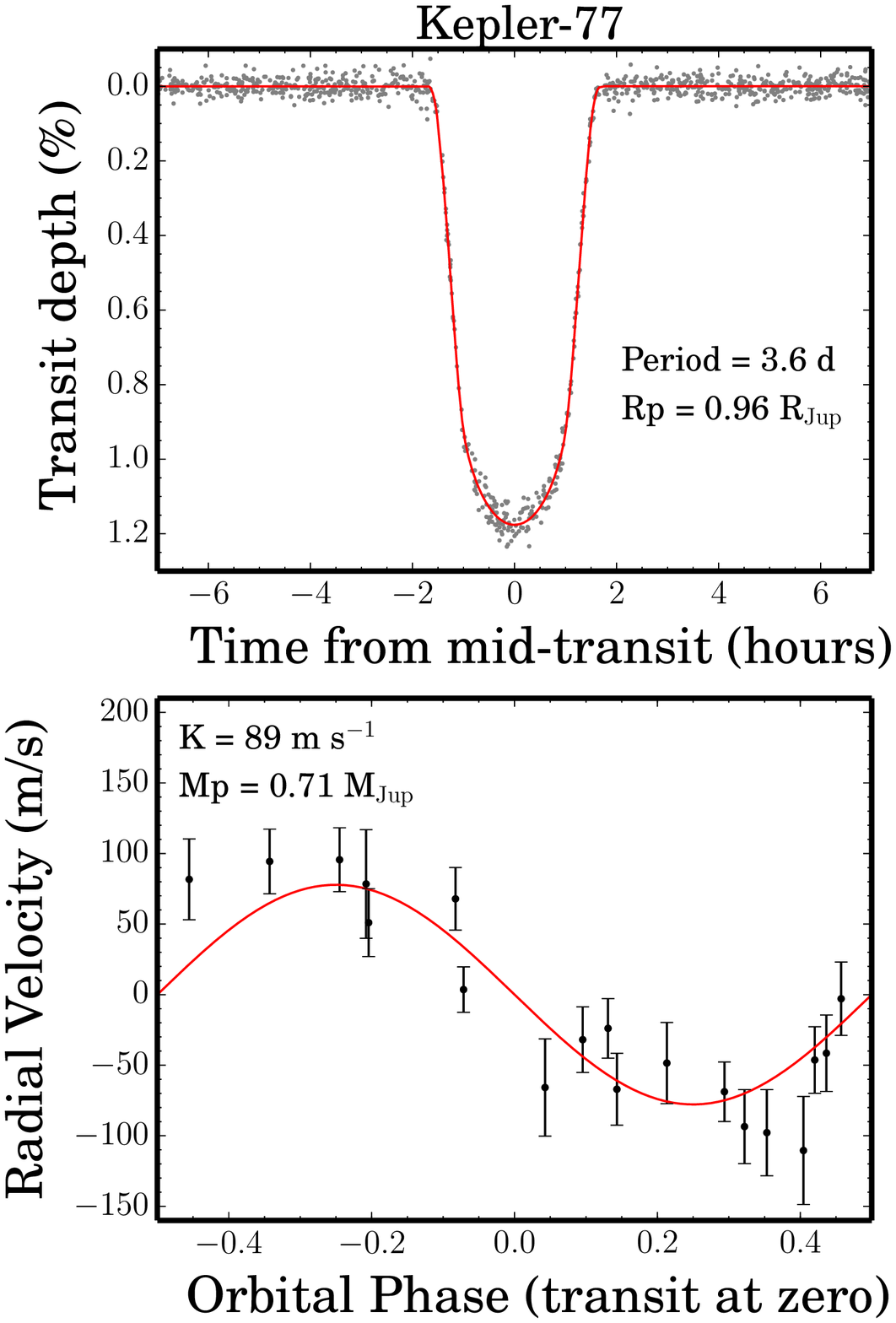}
\caption{Phase-folded {\it Kepler} photometry of Kepler-422 (KOI-22, left) and Kepler-77 (KOI-127, right) along with the best-fit 
transit model from the
MCMC analysis (top panels). The RV data and the best-fit Keplerian orbital solutions are shown
in the bottom panels. 
The RV uncertainties include an additional RV$_{\rm jitter}$ of 22.8\,m\,s$^{-1}$ (Kepler-422) and 14.8\,m\,s$^{-1}$ 
(Kepler-77) from
the highest probability models from the MCMC analysis.
\label{22_127}}
\end{figure}

\begin{figure}
\includegraphics[angle=0,scale=0.55]{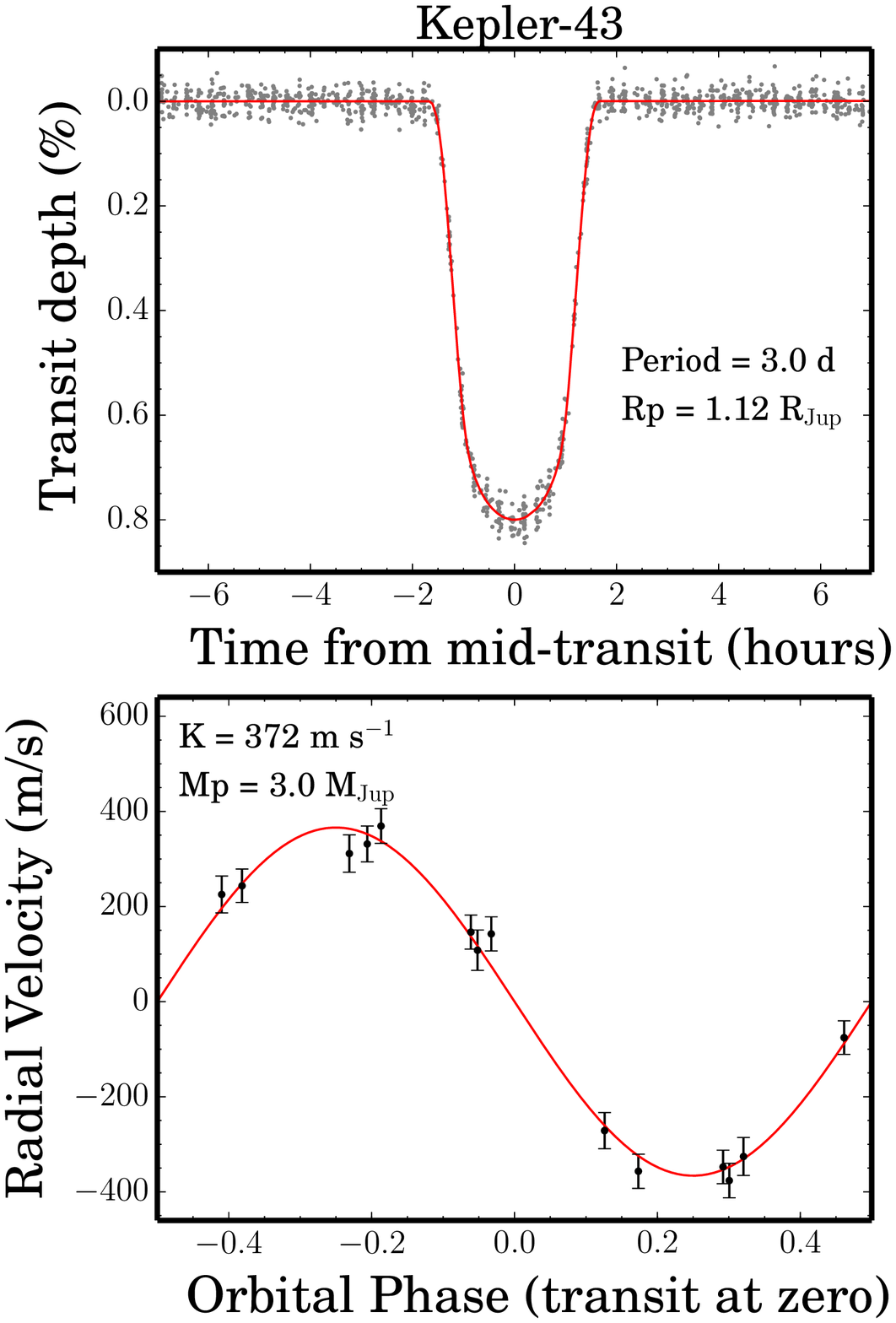}
\includegraphics[angle=0,scale=0.55]{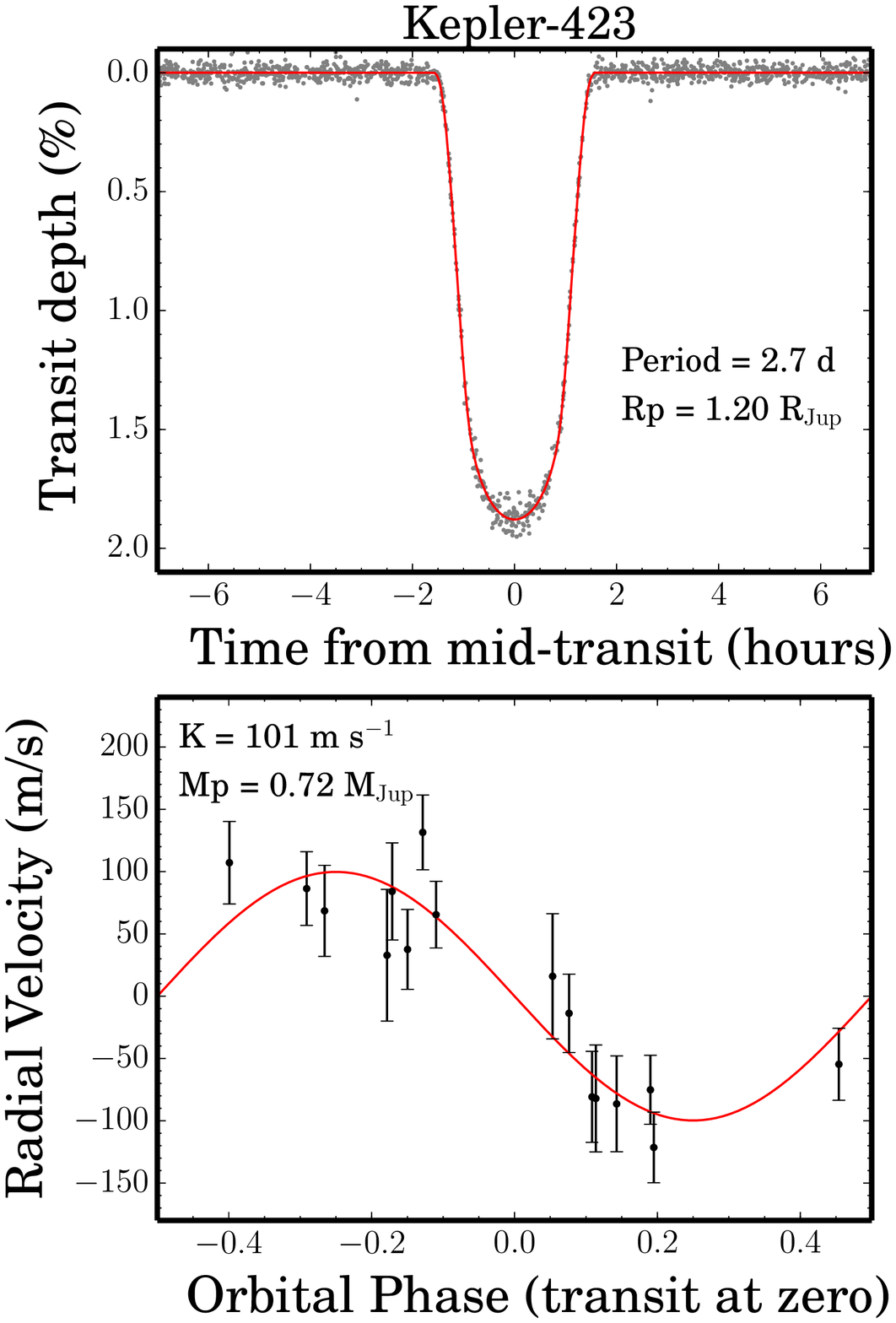}
\caption{Phase-folded {\it Kepler} photometry of Kepler-43 (KOI-135, left) and Kepler-423 (KOI-183, right) along with the best-fit 
transit model from the
MCMC analysis (top panels). The RV data and the best-fit Keplerian orbital solutions are shown
in the bottom panels. The RV uncertainties include an additional RV$_{\rm jitter}$ of 32.7\,m\,s$^{-1}$ (Kepler-43) and 
25.1\,m\,s$^{-1}$
(Kepler-423) from the highest probability models from the MCMC analysis.
\label{135_183}}
\end{figure}

\begin{figure}
\includegraphics[angle=270,scale=0.6]{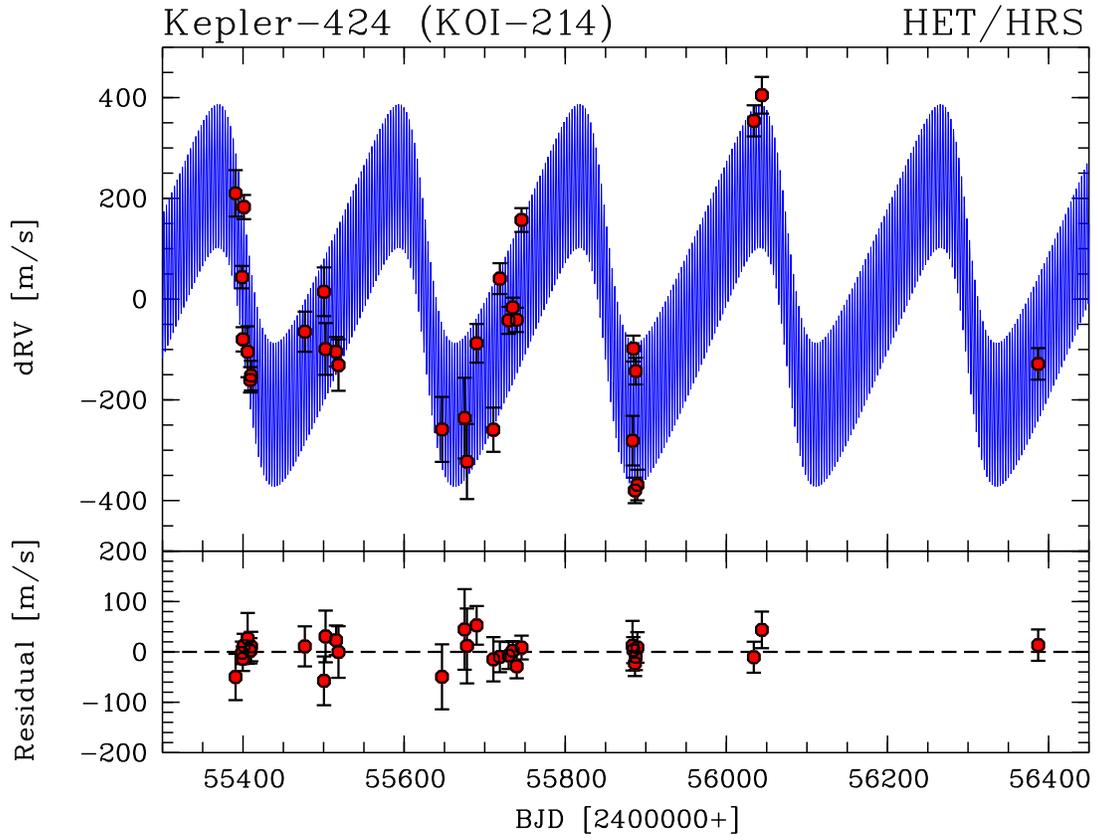}
\caption{{\bf Top panel:} time series of our HET/HRS RV measurements (red points) compared to the 2-planet orbital
solution (solid line). {\bf Lower panel:} RV residuals after subtraction of the orbit, showing no apparent variability 
indicative of additional companions.
\label{koi214time}}
\end{figure}

The resulting planetary, orbital and host star parameters and their uncertainties are summarized in Table\,\ref{214t} (for the 
Kepler-424 two-planet system), 
Table\,\ref{22t} (Kepler-422), Table\,\ref{127t} (Kepler-77), Table\,\ref{135t} (Kepler-43), and Table\,\ref{183t} (Kepler-423). The 
phase-folded 
{\it Kepler} photometry with the best-fit transit model and the RV data along with the best-fit Keplerian orbital solutions are displayed in
Figure\,\ref{214} (Kepler-424), Figure\,\ref{22_127} (Kepler-422 \& Kepler-77) and Figure\,\ref{135_183} (Kepler-43 \& Kepler-423). 
Except in the case of
Kepler-424 we did not find any indication for additional companions in the RV residuals, primarily owing to the small data quantities. 
We performed a period 
search in the RV residuals from the 2-planet model for Kepler-424 (where we have 30 measurements), and did not find any significant 
periodicities. 
Figure\,\ref{koi214time} displays the time series of the RV data along with the 2-planet solution and the residuals from this orbit. 

\section{Planetary evolution models}

\begin{figure}[htbp]
\begin{center}
\includegraphics[scale=0.8]{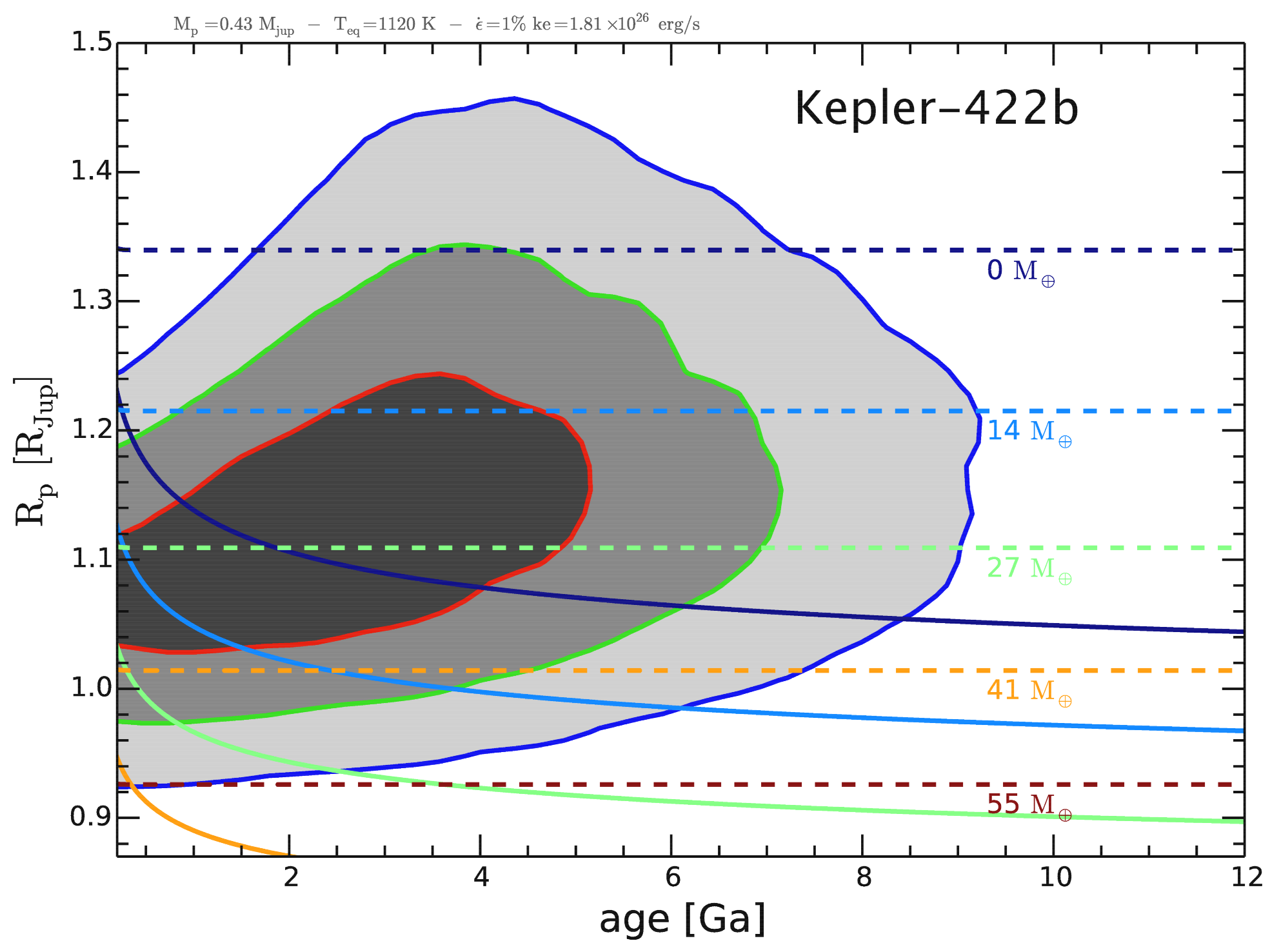}
\caption{Evolution of Kepler-422b's radius as a function of the age. The 68.3\%, 95.5\% and 99.7\% confidence regions are denoted by 
black, dark grey,
and light grey areas respectively.
The curves represent the thermal evolution of a 0.43\,M$_{\rm Jup}$ planet with a time-averaged equilibrium temperature of 1120\,K.
Text labels indicate the amount of heavy elements in the planet (its core mass, in Earth masses). Dashed lines represent planetary evolution
models for which 0.25\% of the incoming stellar flux ($\dot{\epsilon}$) is dissipated into the core of the planet,
whereas plain lines do not account for this dissipation (standard models).}
\label{fig-k22b}
\end{center}
\end{figure}

\begin{figure}[htbp]
\begin{center}
\includegraphics[scale=0.8]{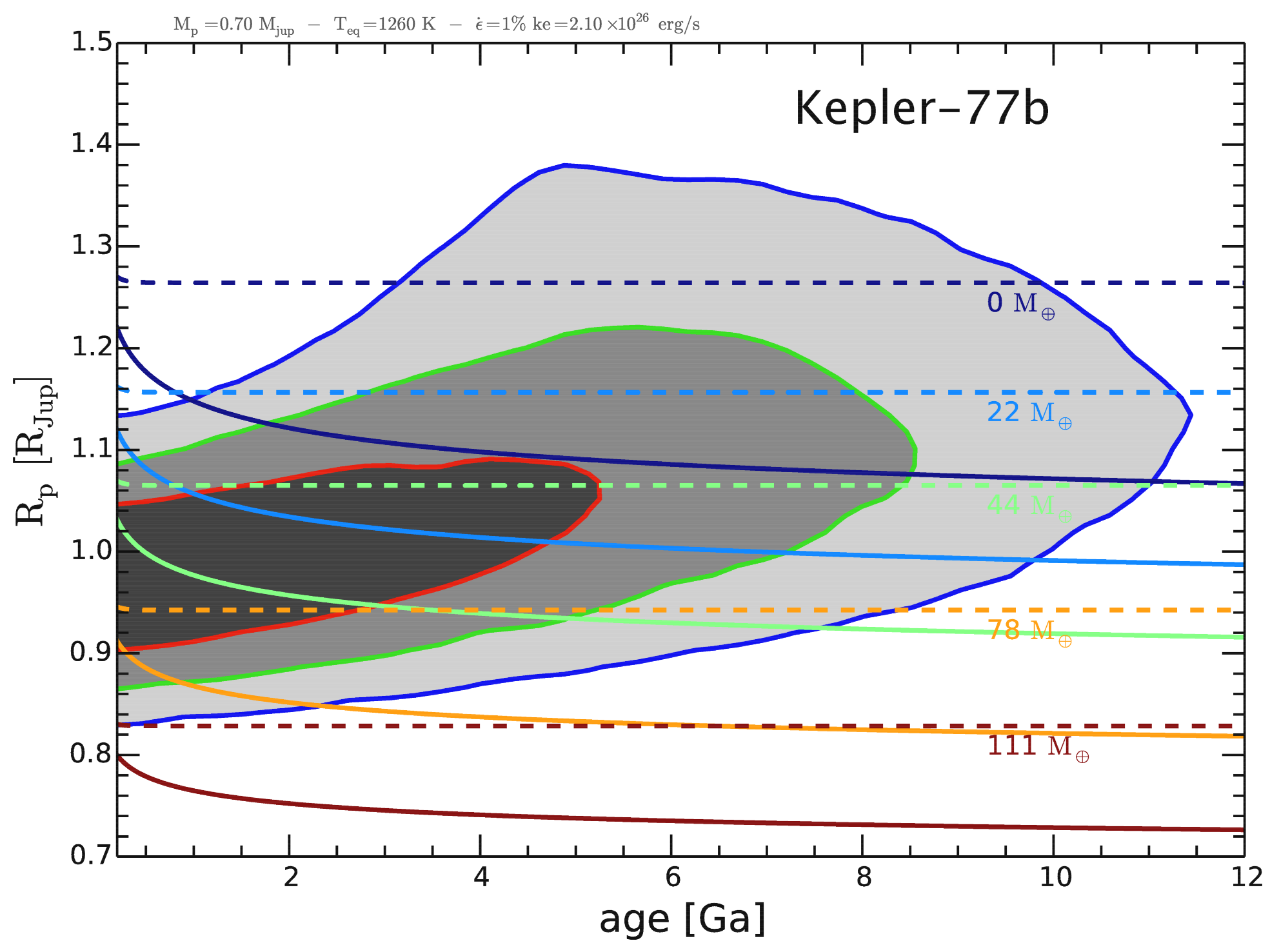}
\caption{Same as Fig. \ref{fig-k22b}, but for Kepler-77b (0.70\,M$_{\rm Jup}$ and 1260\,K)}
\label{fig-k127b}
\end{center}
\end{figure}

\begin{figure}[htbp]
\begin{center}
\includegraphics[scale=0.8]{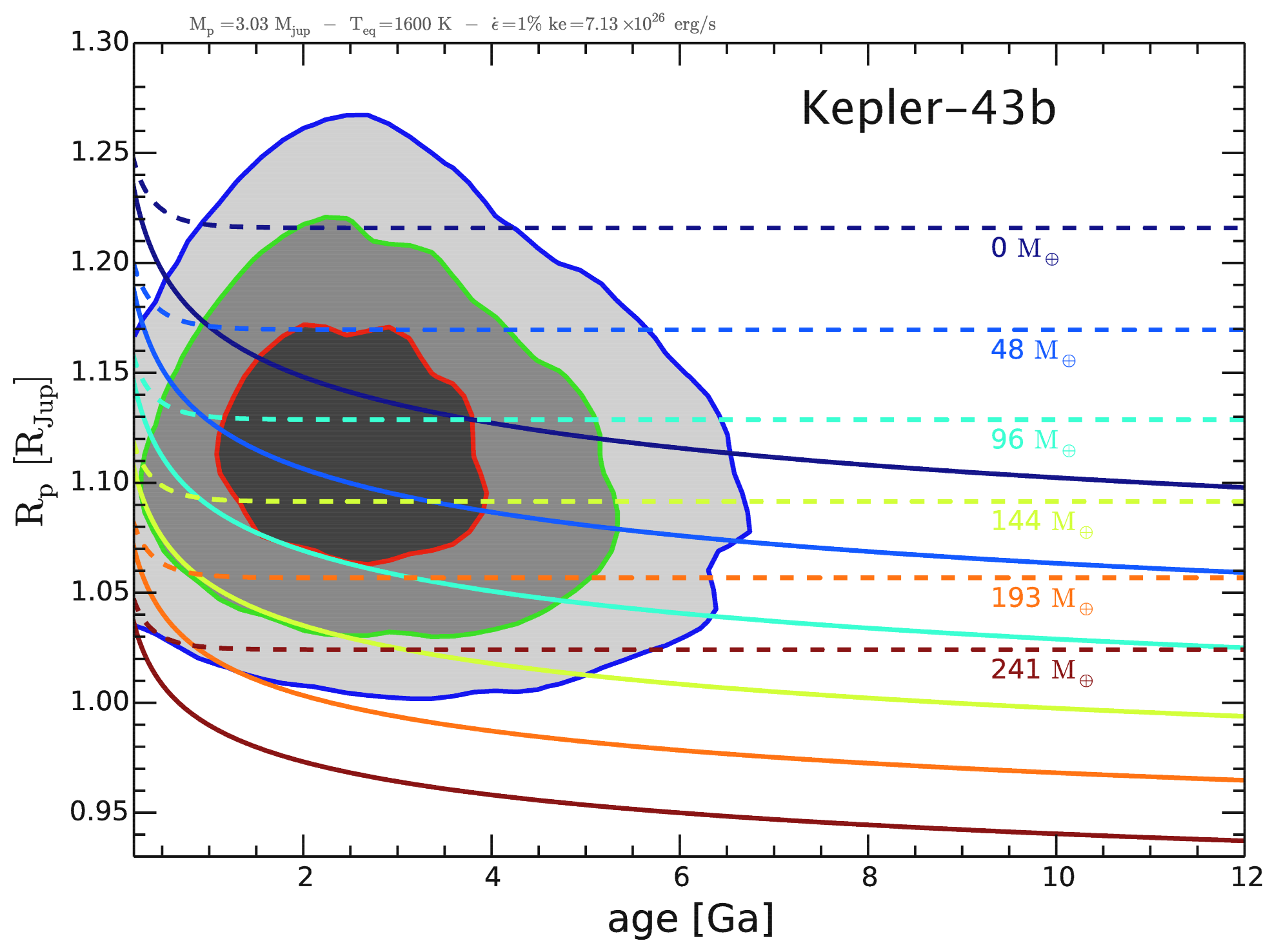}
\caption{Same as Fig. \ref{fig-k22b}, but for Kepler-43b (3.03\,M$_{\rm Jup}$ and 1600\,K)}
\label{fig-k135b}
\end{center}
\end{figure}

\begin{figure}
\begin{center}
\includegraphics[scale=0.8]{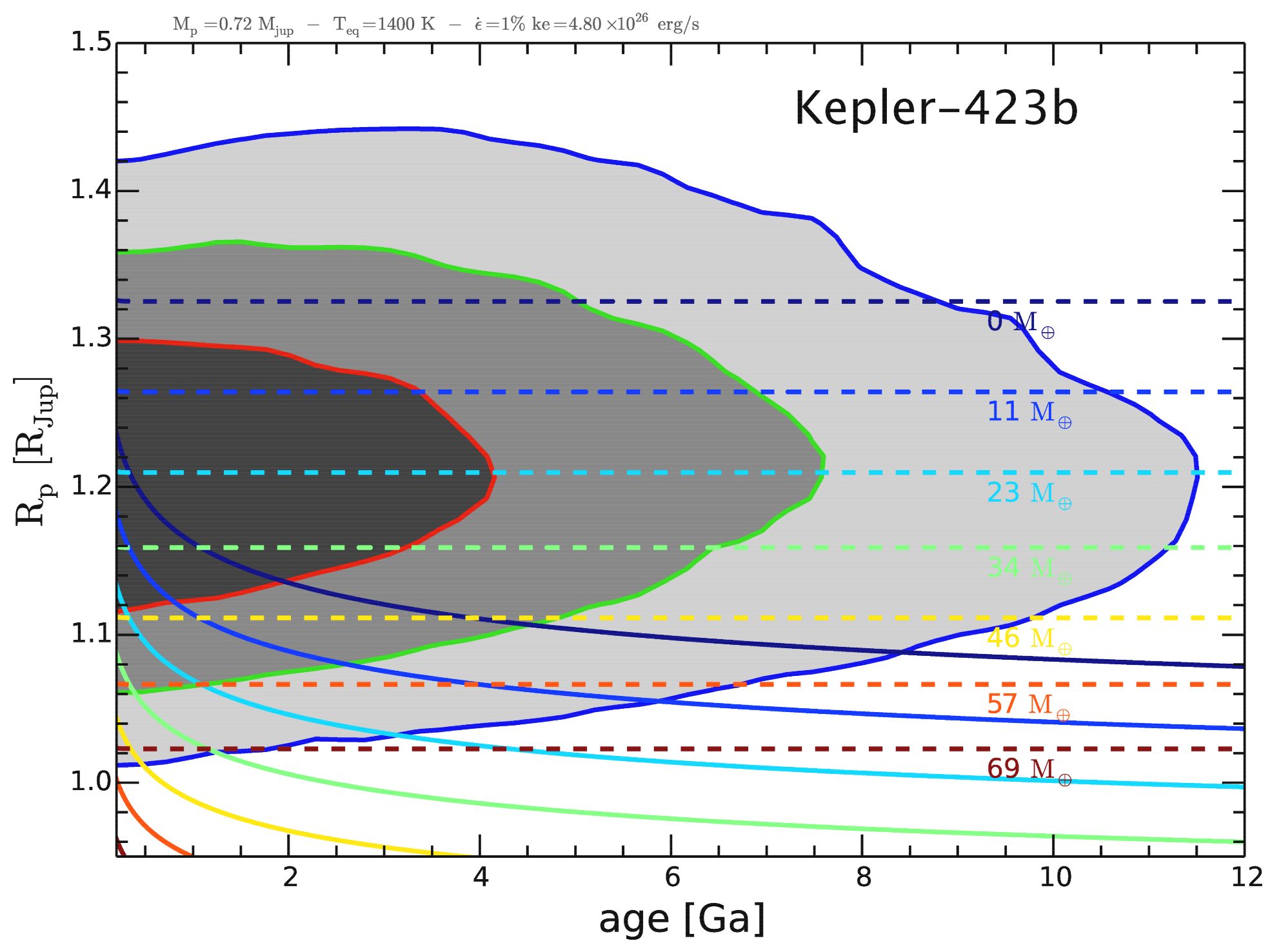}
\caption{Same as Fig. \ref{fig-k22b}, but for Kepler-423b (0.72\,M$_{\rm Jup}$ and 1400\,K)}
\label{fig-k183b}
\end{center}
\end{figure}

\begin{figure}
\begin{center}
\includegraphics[scale=0.8]{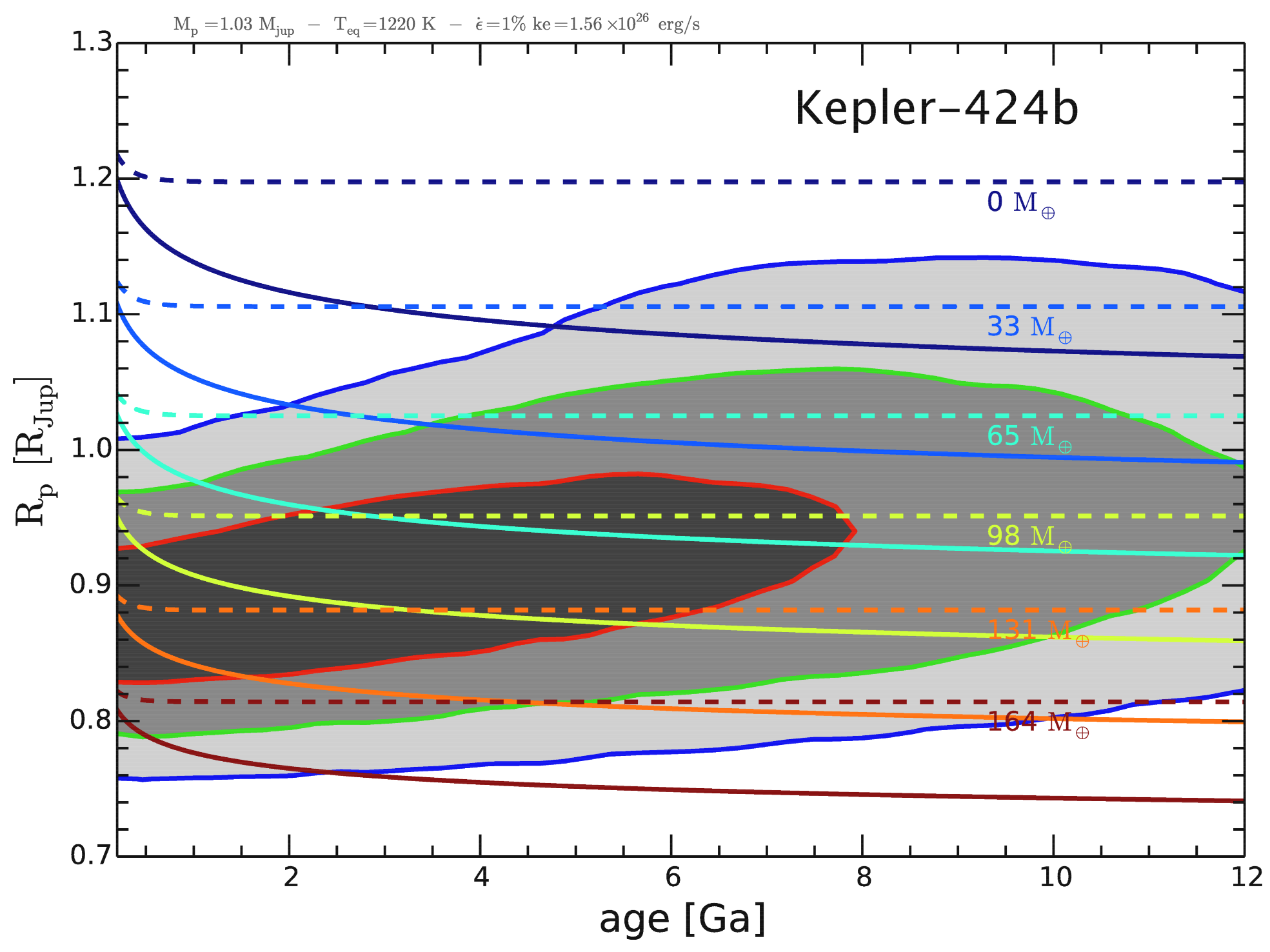}
\caption{Same as Fig. \ref{fig-k22b}, but for Kepler-424b (1.03\,M$_{\rm Jup}$ and 1220\,K)}
\label{fig-k214b}
\end{center}
\end{figure}

With masses between 0.43\,$\rm M_{Jup}$ and 3.03\,$\rm M_{Jup}$, and equilibrium temperature 
above 1100\,K but less than 1600\,K (time-averaged over the orbit), these planets are irradiated giant planets.
We used CEPAM (Guillot \& Morel~1995, Guillot~2010) to build a proper grid of planetary evolution models for each planet (based on their mass, age, 
equilibrium temperature, and unknown heavy elements content). For consistency with previous studies 
(e.g. Miller \& Fortney~2011, Almenara et al.~2013, Deleuil et al.~2014)
models have been calculated in two cases: (i) a ``standard'' case, for which the planet is assumed to be made of a central rocky core surrounded by a 
solar-composition envelope ; (ii) a ``dissipated-energy'' model in which, in addition to the standard case, a fraction (1\%) of the incoming stellar flux 
is assumed to be dissipated in the deep layers of the planet 
(for detailed discussions on possible physical mechanisms, see e.g. Guillot \& Showman~2002, Spiegel \& Burrows~2013, Batygin, Stevenson \& Bodenheimer~2011, Leconte \& 
Chabrier~2012).
In every case, we do not know wether the heavy elements are concentrated in a core, dispersed in the envelope, or a mix of both. 
However, as Baraffe, Chabrier \& Barman~(2008)  
have shown, dispersing heavy elements in the envelope will tend to produce, at a given age, a smaller planet compared to a core-only model. 
Hence our models should provide a lower-limit for the total mass of elements heavier than helium.
Since the absolute planetary parameters fully depend on that of the parent star, and both are model-dependent, we combined stellar 
(PARSEC; Bressan et al.~2012) and planetary evolution models using SET\footnote{Stars \& Exoplanets modeling Tools} 
(Guillot \& Havel~2011, Havel et al.~2011).
Through the use of SET's statistical algorithm and using the observed values only, we thus obtain posterior probability distributions of the 
bulk composition of the planet (i.e. its core mass), as well as independent results for the fundamental parameters of both the star and planet, the latter 
being entirely consistent with those presented in Section\,\ref{mcmc} and Tables\,\ref{214t} to \ref{183t}.
The results are presented in terms of planetary radii as a function of age in \figref{fig-k22b} to \figref{fig-k214b}~: the 68.3, 95.5 and 99.7\% confidence regions 
from the modeling of the star and transit are shown as black, dark grey and light grey area respectively, while the lines show a subset of planetary models 
for the nominal mass and equilibrium temperature of the planet at different compositions as labeled.
With a radius of 1.15\,$\rm R_{Jup}$ and a mass of 0.43\,$\rm M_{Jup}$, Kepler-422b is a slightly inflated Saturn-like planet 
for which the core mass is estimated to be $19^{+20}_{-19}\rm\,M_{\oplus}$\footnote{results from independent 1-D posterior distributions}, or said otherwise 
has a heavy elements mass fraction, $Z$, of $0.14^{+0.15}_{-0.14}$. As \figref{fig-k22b} shows, models without dissipation only cover the lower half of the 
68.3\% confidence region. Although coreless models are possible solutions, Kepler-422b more likely has a significant amount of heavy 
elements because of the high 
metallicity of the parent star (e.g. Guillot et al.~2006, Guillot 2008, Miller \& Fortney~2011). 
Kepler-77b, with a mass of 0.70\,$\rm M_{Jup}$ and a radius of 0.96\,$\rm R_{Jup}$, is a Jupiter-like planet on a slightly eccentric 
orbit ($e \sim 0.23$). 
Its analysis results in a core mass of $48^{+32}_{-44}\rm\,M_{\oplus}$, or equivalently $Z = 0.22^{+0.14}_{-0.20}$. 
Given the age of the system ($2.4^{+2.4}_{-1.6}$ Ga) and the short period (3.6 days), it is surprising that this planet's orbit was able to
maintain such a moderate non-zero eccentricity.
Interestingly, Kepler-423b, despite being a planet of similar mass (0.72\,$\rm M_{Jup}$) around an almost twin star (although much less 
metallic than Kepler-77b), is the 
most inflated planet among these five objects: 1.198\,$\rm R_{Jup}$. This is due to the fact the planet is orbiting closer to its host 
star (2.7 days vs. 3.6 days), and 
therefore has a higher equilibrium temperature (1400\,K vs. 1260\,K). Thus, this is with no surprise that \figref{fig-k183b} shows that standard models cannot 
explain all probable radii solutions (within the 68.3\% confidence region): the planet does have a significant 
positive radius anomaly (see e.g. Laughlin, Crismani \& Adams~2011). 
Its inferred core mass is $25^{+15}_{-10}\rm\,M_{\oplus}$ ($Z = 0.11^{+0.06}_{-0.07}$), although a planet with no core is still possible (but unlikely due to the 
metallicity correlation between host stars and giant planet companions).
Then, there is Kepler-424b, a hot Jupiter of mass 1.03\,$\rm M_{Jup}$ and radius 0.886\,$\rm R_{Jup}$. 
Its derived core mass is $119^{+35}_{-27}\rm\,M_{\oplus}$ ($Z = 0.36^{+0.11}_{-0.08}$). The system has also a second giant planet, 
which is not transiting, and therefore we cannot have refined estimates on the 
bulk composition of Kepler-424b like in the case of Kepler-9 (Havel et al.~2011).
Anyway, Kepler-424b appears to require a large fraction of heavy elements (as compared to Jupiter) in order to 
explain its observed radius.
And finally, the most massive planet of the sample, Kepler-43b (3.03\,$\rm M_{Jup}$ and 1.124\,$\rm R_{Jup}$) has a 
core mass of $74^{+83}_{-32}\rm\,M_{\oplus}$ ($Z = 0.08^{+0.08}_{-0.04}$). Compared to the other planets of this paper, Kepler-43b 
really is the 
most inflated planet (not in absolute size, but relative to its mass). 
In fact it has the highest equilibrium temperature of the five planets (1600\,K), and 
more notably has the best age constraint (due to a more precise stellar density measurement from the transit).
Therefore, the five transiting planets presented in this paper have heavy elements content ranging from 20 to 120 M$_\oplus$. 
At least three of them may be coreless (Kepler-422b, Kepler-43b, Kepler-423b ; Kepler-77b is on the edge), and only one definitely 
requires a 
non-zero metallicity (Kepler-424b, part of a multiple system). Lastly, errors on the given core masses may be larger due to 
non-considered 
uncertainties in the physics of the planetary models (equation of state, opacities, dissipation, ... ; see e.g. Vazan et al.~2013, Militzer \& Hubbard~2013, Spiegel \& 
Burrows~2013), 
which could increase the modeled radii by up to $\sim$ 50\%, and therefore would result in larger (heavier) cores.

\section{Discussion}

Two of the hot Jupiters presented here, KOI-127.01 and KOI-135.01, have been confirmed previously by other groups in the
astronomical community. 

KOI-127.01 was confirmed as Kepler-77b by Gandolfi et al.~(2013).
Their spectroscopic data were obtained with the Sandiford
spectrograph at the Otto Struve 2.1\,m telescope at McDonald Observatory and with FIES at the Nordic Optical Telescope on La Palma.
In general, their results are in good agreement with our work. One discrepancy is the host star metallicity, where we
find [Fe/H]$=0.43\pm0.1$ while Gandolfi et al. determined an [M/H] of $0.20\pm0.05$. They also estimated a somewhat cooler
effective temperature of $5520\pm60$\,K, which is almost consistent with our value of $5668\pm77$\,K. The largest difference
is the planetary mass (and thus also planet density): they obtain $0.43\pm0.032~M_{\rm Jup}$ compared to our value of
$0.7\pm0.1~M_{\rm Jup}$. This discrepancy is due to a smaller RV semi-amplitude of $K=59.2\pm4.3$\,m\,s$^{-1}$ measured from the
Sandiford and FIES RVs. In contrast, the $K$ value that we have determined from our HRS and FIES RVs is $K=89\pm11$\,m\,s$^{-1}$.

KOI-135.01 was confirmed as Kepler-43b by Bonomo et al.~(2012) using spectroscopic observations with the SOPHIE spectrograph on
the 1.93\,m telescope at the Observatoire de Haute Provence. In this case, the results from both of our studies are fully consistent
within the quoted uncertainties.

Of these five systems, the Kepler-424 system stands out as a multi-planet system that also contains a hot Jupiter.
The second companion resides on a moderately eccentric orbit and is many times more massive than the inner planet (mass ratio of $\sim 7:1$).    
Given the large separation between the two planets in the Kepler-424 system, intuition suggests their orbits should be dynamically 
stable over long 
timescales.  However, as evidenced by the $\upsilon$ And system (McArthur et al.~2010), giant planets with mutually inclined orbital planes may exhibit 
important dynamical interactions, even when the orbits are well separated. To test whether this is the case for Kepler-424, we have 
performed some very simple dynamical simulations of the system using the SYSTEMIC console (Meschiari et al.~2009).

In our simulations, the planets are initially configured with the orbital solution given in Table\,\ref{214t}.
We left the inclination of planet b fixed, while setting c to a range of mutual inclinations in steps of 5 degrees.
At each mutual inclination we simulated the orbits over 100,000 years, or until a collision occurred.
The simulations were computed using a Gragg-Burlisch-St\"oer integrator with a time step of 1 day.

\begin{figure}
\includegraphics[angle=0,scale=0.65]{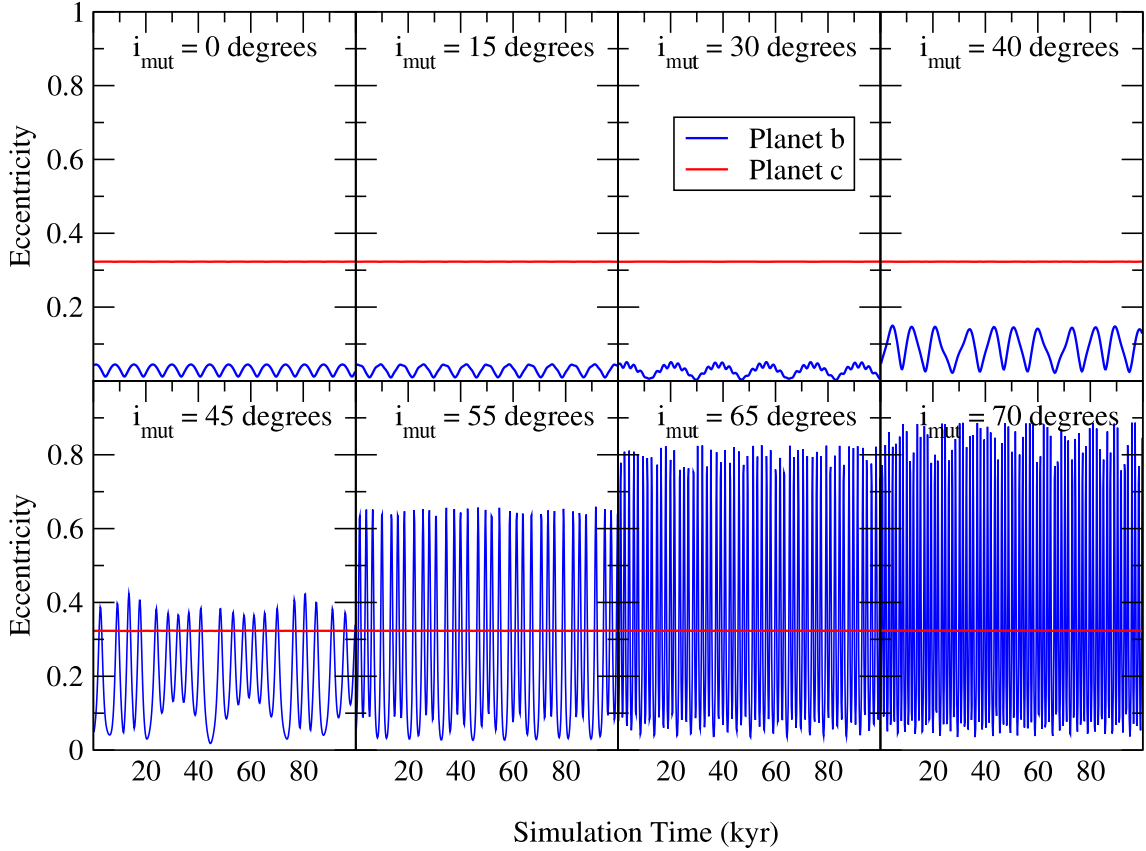}
\caption{Dynamical simulations of the Kepler-424 two-planet system showing eccentricity as a function of time. 
Each panel shows the evolution of the planets' eccentricities for a different mutual inclination $i_{mut}$ of the planetary orbits.
The high eccentricities of the orbit of planet b induced by $i_{mut} \geq 45^{\circ}$ are inconsistent with the 
observed low or zero eccentricity of the hot Jupiter orbit. This would correspond to an 
estimated upper mass limit for planet c of $M_{\rm c} \leq 9.6~M_{\rm Jup}$.  
\label{paul}}
\end{figure}

The effect of planet c's inclination on the system stability can best be visualized by its impact on the eccentricity of planet b.
In Figure\,\ref{paul}, we show the planets' eccentricities as a function of simulation time for a representative set of mutual inclination values.
Even for coplanar orbits, the eccentricity of b undergoes low-amplitude periodic oscillations, but the amplitude of the oscillations increases
greatly for higher mutual inclinations.  Most notably, when the mutual inclination reaches 45 degrees, $e_\textrm{b}$ abruptly increases to
levels inconsistent with our orbital solution. This sudden increase of planet b's eccentricity is caused by the Lidov-Kozai mechanism (Lidov 
1962, Kozai 1962). While it is conceivable that the system is highly mutually inclined and we are observing it during a period of low 
eccentricity,
this possibility becomes less likely as the inclination of planet c decreases and the mean eccentricity of planet b increases.
We therefore conclude that the inclination of planet c is most likely greater than 45 degrees.

The results of our dynamical analysis are particularly useful because they allow us to place an upper limit on the mass of planet c.
If we require $i_{\textrm{c}} \geq 45^{\circ}$, then the actual mass of planet c is $M_{\rm c} \leq 9.6 M_{\rm Jup}$.
This limit places planet c comfortably within the planetary mass regime, ruling out the possibility that it is a stellar or brown dwarf companion in a face-on 
configuration. A more sophisticated dynamical study of this system, that also includes tidal dissipation, could lead to a more robust upper mass limit for 
planet c.

Observing the Rossiter-McLaughlin effect for Kepler-424b will be an interesting test. If planet-planet scattering is the physical 
mechanism that caused the 
inward migration, it could also lead to a high inclination of the planet's orbit with respect to the stellar equator.  It is a challenging observation for
a $V=14.5$~magnitude star, that we plan to carry out with the newly upgraded and high-efficiency HRS at the HET, once it will become available
in early 2015.

\acknowledgments
Funding for this Discovery mission is provided by NASA's Science Mission Directorate.
We thank the entire {\it Kepler} team, a wonderful group of people who make this mission so 
successful. We also thank the anonymous referee, her/his comments helped to improve this
paper. ME and DH acknowledge support by NASA under Grants 
NNX14AB86G (ME) and NNX14AB92G (DH) issued through the Kepler 
Participating Scientist Program. We are grateful to Geoff Marcy and Andrew Howard for their help in 
obtaining the Keck/HIRES observations. 
The Hobby-Eberly Telescope (HET) is a joint project of the University of
Texas at Austin, the Pennsylvania State University, Stanford University,
Ludwig-Maximilians-Universit\"{a}t M\"{u}nchen,
and Georg-August-Universit\"{a}t G\"{o}ttingen.
The HET is named in honor of its principal benefactors,
William P. Hobby and Robert E. Eberly. 
Based in part on observations made with the Nordic Optical
Telescope, operated on the island of La Palma jointly by
Denmark, Finland, Iceland, Norway, and Sweden, in the Spanish
Observatorio del Roque de los Muchachos of the Instituto de
Astrofisica de Canarias. Some of the data presented herein were obtained at the W.M. Keck Observatory, which is operated as a scientific partnership among the 
California Institute of Technology, the University of California and the National Aeronautics and Space Administration. The Observatory was made possible by 
the generous financial support of the W.M. Keck Foundation. The authors wish to recognize and acknowledge the very significant cultural role and reverence that 
the summit of Mauna Kea has always had within the indigenous Hawaiian community.  We are most fortunate to have the opportunity to conduct observations from 
this mountain. 


\begin{deluxetable}{lrrr}
\tablecolumns{4}
\tablewidth{0pt}
\tablecaption{Precise radial velocity measurements of Kepler-424 (KOI-214)
\label{rv214}}
\tablehead{
\colhead{BJD [days]} & {dRV [m\,s$^{-1}$]} & {error [m\,s$^{-1}$]} & {instrument}
} 
\startdata
2455390.7061  &    -11651.2  &      46.0 & HRS/HET \\
2455398.6922  &        -11817.6  &      22.2 & HRS/HET \\
2455399.6872  &        -11941.0 &       24.2 & HRS/HET \\
2455400.8879  &    -11678.4   &     24.0 & HRS/HET \\
2455405.6585  &    -11965.6   &     50.5 & HRS/HET \\
2455408.9012  &        -12021.4  &      25.1 & HRS/HET \\
2455409.8791  &        -12013.0 &       29.3 & HRS/HET \\
2455476.7011  &    -11925.9  &      39.7 & HRS/HET \\
2455502.6109  &    -11960.1 &    51.3 & HRS/HET \\
2455515.5966  &        -11965.6  &     29.4 & HRS/HET \\
2455518.5628  &         -11992.1 &       51.2 & HRS/HET \\
2455674.9104  &    -12097.2   &     80.0 & HRS/HET \\
2455677.9101  &    -12183.7   &     74.5 & HRS/HET \\
2455689.8917  &    -11948.8   &     38.5 & HRS/HET \\
2455710.8406  &    -12120.4   &     44.0 & HRS/HET \\
2455718.8050  &    -11820.5    &    30.6 & HRS/HET \\
2455729.7733  &    -11903.3   &     26.8 & HRS/HET \\
2455734.7532  &    -11877.7  &      19.5 & HRS/HET \\
2455739.7648  &    -11902.4 &       24.0 & HRS/HET \\
2455745.7485  &        -11704.1     &   23.7 & HRS/HET \\
2455883.5846  &        -12142.2    &    49.3 & HRS/HET \\
2455884.5760  &        -11959.3   &     25.8 & HRS/HET \\
2455886.5792  &        -12240.8  &     25.3 & HRS/HET \\
2455887.5533  &        -12004.1 &       26.5 & HRS/HET \\
2455889.5618  &    -12230.1    &    30.4 & HRS/HET \\
2456033.9355  &       -11507.4  &      30.8 & HRS/HET \\
2456043.9233  &    -11456.5     &   36.4 & HRS/HET \\
\enddata
\end{deluxetable}

\begin{deluxetable}{lrrr}
\tablecolumns{4}
\tablewidth{0pt}
\tablecaption{Precise radial velocity measurements of Kepler-422 (KOI-22)
\label{rv22}}
\tablehead{
\colhead{BJD [days]} & {dRV [m\,s$^{-1}$]} & {error [m\,s$^{-1}$]} & {instrument}
}
\startdata
2455014.9026 &   13.9  &  17.7 & Keck/HIRES \\
2455017.0576 &   15.8  &  17.7 & Keck/HIRES \\
2455041.9829 &  -22.4  &  18.2 & Keck/HIRES \\
2455042.8107 &  -30.7  &  18.0 & Keck/HIRES \\
2455044.0113 &  -54.7  &  18.2 & Keck/HIRES \\
2455045.0133 &  -27.9  &  18.0 & Keck/HIRES \\
2455048.8795 &   85.8  &  18.8 & Keck/HIRES \\
2455074.8224 &  -20.0  &  17.4 & Keck/HIRES \\
2455075.8558 &   -6.9  &  17.4 & Keck/HIRES \\
2455078.8808 &  108.6  &  17.6 & Keck/HIRES \\
2455084.8208 &  -34.0  &  17.4 & Keck/HIRES \\
2455437.8116 &  -28.2  &   4.0 & Keck/HIRES \\
2455465.8030 &   18.6  &   4.0 & Keck/HIRES \\
2455789.8189 &   20.2  &   3.6 & Keck/HIRES \\
2455792.9381 &  -37.2  &   3.8 & Keck/HIRES \\
\enddata
\end{deluxetable}

\begin{deluxetable}{lrrr}
\tablecolumns{4}
\tablewidth{0pt}
\tablecaption{Precise radial velocity measurements of Kepler-77 (KOI-127)
\label{rv127}}
\tablehead{
\colhead{BJD [days]} & {dRV [m\,s$^{-1}$]} & {error [m\,s$^{-1}$]} & {instrument}
}
\startdata
2455432.7847  &   -43.2 &  20.8 & HET/HRS \\
2455442.7527  &    27.5 &   6.4 & HET/HRS \\
2455476.6646  &   -86.6 &  35.3 & HET/HRS \\
2455481.6440  &    74.9 &  19.0 & HET/HRS \\
2455483.6388  &   -73.9 &  26.7 & HET/HRS \\
2455507.5791  &   -41.9 &  31.1 & HET/HRS \\
2455508.5783  &   -69.6 &  21.7 & HET/HRS \\
2455512.5673  &   -17.6 &  22.7 & HET/HRS \\
2455674.8853  &   102.3 &  35.5 & HET/HRS \\
\hline
2455379.6416   &   -44.9  &     15.1 & NOT/FIES \\
2455380.5392   &   105.6  &     24.6 & NOT/FIES \\
2455382.5104   &    -8.0  &     18.0 & NOT/FIES \\ 
2455383.6729   &   -22.4  &     18.4 & NOT/FIES \\
2455384.5210   &   118.3  &     17.5 & NOT/FIES \\
2455428.3972   &    91.8  &     16.6 & NOT/FIES \\
2455429.4562   &   -24.6  &     24.7 & NOT/FIES \\
2455431.3953   &   119.5  &     17.2 & NOT/FIES \\
2455480.4322   &    21.0  &     21.4 & NOT/FIES \\
2455486.4220   &     0.0  &     15.1 & NOT/FIES \\
\enddata
\end{deluxetable}

\begin{deluxetable}{lrrr}
\tablecolumns{4}
\tablewidth{0pt}
\tablecaption{Precise radial velocity measurements of Kepler-43 (KOI-135)
\label{rv135}}
\tablehead{
\colhead{BJD [days]} & {dRV [m\,s$^{-1}$]} & {error [m\,s$^{-1}$]} & {instrument}
}
\startdata
2455382.6432  &     -3.3   &    14.7  & NOT/FIES \\
2455383.6244   &  -493.3   &    13.0 & NOT/FIES \\
2455384.6123  &     97.8   &    12.9 & NOT/FIES \\
2455424.5139  &    223.4   &    15.8 & NOT/FIES \\
2455425.6033  &   -502.4   &    14.9 & NOT/FIES \\
2455426.4740  &   -221.5   &    13.3 & NOT/FIES \\
2455427.4796  &    185.7   &    18.9 & NOT/FIES \\
2455428.4842  &   -417.0   &    19.3 & NOT/FIES \\
2455430.4280  &    165.6   &    21.7 & NOT/FIES \\
2455480.4818  &   -471.4   &    23.0 & NOT/FIES \\
2455482.3520  &      0.4   &    14.5 & NOT/FIES \\
2455486.4697  &   -522.0   &    15.8 & NOT/FIES \\
2455488.4282  &    -37.7   &    27.1 & NOT/FIES \\
2455490.3694  &     79.6   &    20.8 & NOT/FIES \\
\enddata
\end{deluxetable}

\newpage

\begin{deluxetable}{lrrr}
\tablecolumns{4}
\tablewidth{0pt}
\tablecaption{Precise radial velocity measurements of Kepler-423 (KOI-183)
\label{rv183}}
\tablehead{
\colhead{BJD [days]} & {dRV [m\,s$^{-1}$]} & {error [m\,s$^{-1}$]} & {instrument}
}
\startdata
      2455398.6748  &    30.5 &  43.5 & HET/HRS \\
      2455400.8711  &   146.0 &  16.4 & HET/HRS \\
      2455405.8708  &    83.0 &  26.6  & HET/HRS \\
      2455448.7528  &   100.9 &  15.6  & HET/HRS \\
      2455451.7393  &    47.4 &  46.5  & HET/HRS \\
      2455452.7287  &   -60.7 &  11.7  & HET/HRS \\
      2455476.6827  &   -67.6 &  34.8  & HET/HRS \\
      2455487.6378  &  -106.9 &  13.2  & HET/HRS \\
      2455511.5646  &   -66.3 &  26.5  & HET/HRS \\
      2455513.5558  &    52.1 &  20.0  & HET/HRS \\
      2455515.5707  &   121.7 &  21.5  & HET/HRS \\
      2455520.5452  &   -40.2 &  14.4  & HET/HRS \\
      2455521.5509  &    98.6  & 29.8  & HET/HRS \\
      2456088.7878  &   -71.9  & 29.2  & HET/HRS \\
      2456104.7149  &     0.8  & 18.9  & HET/HRS \\
      2456219.6414  &    80.0  &  9.1  & HET/HRS \\
\enddata
\end{deluxetable}

\begin{deluxetable}{lrrrrr}
\tablecolumns{6}
\tablewidth{0pt}
\tablecaption{Basic parameters of the target stars: Kepler, KOI, and KeplerID numbers are given, along
with {\it Kepler} magnitude $K_{\rm p}$, $V$ band magnitude and coordinates.
\label{tab1}}
\tablehead{
\colhead{KOI} & {KepID} & {$K_{\rm p}$[mag]} & {$V$[mag]} & {RA [2000]} & {DEC [2000]}
}
\startdata
22 & 9631995 & 13.435 & 13.642   & 18:50:31.11 & 46:19:24.10 \\
127 & 8359498 & 13.938 & 14.197  & 19:18:25.92 & 44:20:43.52 \\
135 & 9818381 & 13.958 & 14.082 & 19:00:57.78 & 46:40:05.70 \\
183 & 9651668 & 14.290 & 14.499  & 19:31:25.36 & 46:23:28.24 \\
214 & 11046458 & 14.256 & 14.497 & 19:54:29.99 & 48:34:38.82 \\
\enddata
\end{deluxetable}

\begin{deluxetable}{lrrrrrl}
\tablecolumns{7}
\tablewidth{0pt}
\tablecaption{Spectroscopic stellar parameters of the five target stars. Note that, following Torres et al.~(2012), we
added in quadrature additional uncertainties of 59~K in T$_{\rm eff}$, 0.062 dex in [Fe/H] and 
0.85 km\,s$^{-1}$ in $v$ sin$i$ to account for systematic uncertainties.  
\label{stars}}
\tablehead{\colhead{KOI} & {T$_{\rm eff}$[K]} & {log g} & {$v$ sin$i$ [km/s]} & {[Fe/H]} & {[M/H]} & {notes}}
\startdata
  22 & 5972$\pm$84 & 4.50$\pm0.1$ & 2.2$\pm$1.3 & 0.23$\pm$0.09 & ... & SME\\
 127 & 5668$\pm$77 & 4.53$\pm0.1$ & 1.8$\pm$1.0 & ... & 0.43$\pm$0.1 & SPC \\
 135 & 6019$\pm$82 & 4.54$\pm0.1$ & 4.9$\pm$1.0 & ... & 0.43$\pm$0.1 & SPC  \\
 183 & 5790$\pm$116 & 4.57$\pm0.12$ & ... & 0.26$\pm$0.12 & ... & MOOG\\
 214 & 5460$\pm$81 & 4.49$\pm0.05$ & 0.5$\pm$1.0 & 0.44$\pm$0.14 & ... & MOOG/SPC/SME\\
\enddata
\end{deluxetable}

\begin{deluxetable}{lrrrr}
\tablecolumns{4}
\tablewidth{0pt}
\tablecaption{Parameters of the Kepler-424 (KOI-214) planetary system
\label{214t}}
\tablehead{
\colhead{Parameter [unit]} & {median} & {+1$\sigma$} & {-1$\sigma$} & 
{notes}
}
\startdata
$M_{\star}$ [$M_{\odot}$] & 1.01 & +0.054 & -0.054 & isochrone fit\\
$R_{\star}$ [$R_{\odot}$] & 0.94 & +0.056 & -0.056 & isochrone fit\\
log(g) & 4.50 & +0.05 & -0.05 & isochrone fit\\
$L_{\star}$ [$L_{\odot}$] & 0.71 & +0.11 & -0.11 & isochrone fit \\ 
$\rho_{\star}$ [g\,cm$^{-3}$] & 1.73 & +0.29 & -0.29 & isochrone fit\\
$\rho_{\star}$ [g\,cm$^{-3}$] & 1.74 & +0.44 & -0.34 & transit model\\
\hline
Planet b: & & & & \\
\hline
$P$ [days]& 3.3118644 & $3.9\times10^{-7}$ & -3.9$\times10^{-7}$ & \\
$T0$ [BJD] & 2454964.7427 & +0.00023 & -0.00017 & \\
$b$ & 0.934 & +0.065 & -0.053 & \\
$R_{\rm planet} / R_{\star}$ & 0.0961 & +0.0065 & -0.0033 &\\
$R$ [$R_{\rm Jup}$] & 0.89 & +0.08 & -0.06 &\\
$K$ [m\,s$^{-1}$] & 140.0 & +12.0 & -13.0 & \\
$e \cos \omega$ & 0.001 & +0.043 & -0.029 & \\
$e \sin \omega$ & 0.002 & +0.061 & -0.066 & \\
$M$ [$M_{\rm Jup}$] & 1.03 & +0.13 & -0.13 & \\
$a$ [AU] & 0.044 & +0.005 & -0.004 & \\
$\rho$ [g\,cm$^{-3}$] & 1.94 & +0.25 & -0.25 & \\
\hline
Planet c: & & & & \\
\hline
$P$ [days]& 223.3 & +2.1 & -2.1 & \\
$T0$ [BJD] & 2455403.4  & +2.1 & -2.0 &\\
$K$ [m\,s$^{-1}$] & 246.0 & +17.0 & -17.0 & \\
$e \cos \omega$ & 0.018 & +0.052 & -0.052 & \\
$e \sin \omega$ & 0.318 & +0.057 & -0.061 & \\
$M \sin i$ [$M_{\rm Jup}$] & 6.97 & +0.62 & -0.62 &\\
$a$ [AU] & 0.73 & +0.08 & -0.07 &\\
\hline
RV$_{\rm jitter}$ [m\,s$^{-1}$] & 8.0 & +13.0 & -6.0 & \\
\hline
\enddata
\end{deluxetable}

\begin{deluxetable}{lrrrr}
\tablecolumns{4}
\tablewidth{0pt}
\tablecaption{Parameters of the Kepler-422 (KOI-22) transiting system
\label{22t}}
\tablehead{
\colhead{Parameter [unit]} & {median} & {+1$\sigma$} & {-1$\sigma$} & notes
}
\startdata
$M_{\star}$ [$M_{\odot}$] & 1.15 & +0.06 & -0.06 & isochrone fit\\
$R_{\star}$ [$R_{\odot}$] & 1.24 & +0.12 & -0.12 & isochrone fit\\
log(g) & 4.31 & +0.073 & -0.073 & isochrone fit\\
$L_{\star}$ [$L_{\odot}$] & 1.75 & +0.37  & -0.37 & isochrone fit\\
$\rho_{\star}$ [g\,cm$^{-3}$] & 0.86 & +0.22 & -0.22 & isochrone fit\\
$\rho_{\star}$ [g\,cm$^{-3}$] & 0.87 & +0.22 & -0.22 & transit model \\
\hline
Planet : & & & & \\
\hline
$P$ [days]& 7.8914483 & +5.0$\times 10^{-7}$ & -5.1$\times 10^{-7}$ &  \\
$T0$ [BJD] & 2455010.25005 & +0.00011 & -9.4$\times 10^{-5}$ & \\
$b$ & 0.416 & +0.045 & -0.045 &  \\
$R_{\rm planet} / R_{\star}$ & 0.0957 & +0.00048 & -0.00055 &  \\
$R$ [$R_{\rm Jup}$] & 1.15 & +0.11 & -0.11 & \\
$K$ [m\,s$^{-1}$] & 40.0 & +11.0 & -10.0 & \\
$e \cos \omega$ & 0.013 & +0.096 & -0.063 & \\
$e \sin \omega$ & -0.009 & +0.07 & -0.096 & \\
$M$ [$M_{\rm Jup}$] & 0.43 & +0.13 & -0.13 & \\
$a$ [AU] & 0.082 & +0.011 & -0.010  & \\
$\rho$ [g\,cm$^{-3}$] & 0.38 & +0.11 & -0.11 & \\
\hline
RV$_{\rm jitter}$ [m\,s$^{-1}$] & 26.8 & +9.4 & -7.0 & \\
\hline
\enddata
\end{deluxetable}

\begin{deluxetable}{lrrrr}
\tablecolumns{5}
\tablewidth{0pt}
\tablecaption{Parameters of the Kepler-77 (KOI-127) transiting system
\label{127t}}
\tablehead{
\colhead{Parameter [unit]} & {median} & {+1$\sigma$} & {-1$\sigma$} & {notes}
}
\startdata
$M_{\star}$ [$M_{\odot}$] & 1.08 & +0.034 & -0.034 & isochrone fit\\
$R_{\star}$ [$R_{\odot}$] & 0.99 & +0.053 & -0.053 & isochrone fit\\
log(g) & 4.48 & +0.036 & -0.036 & isochrone fit\\
$L_{\star}$ [$L_{\odot}$] & 0.90 & +0.13 & -0.13 & isochrone fit \\
$\rho_{\star}$ [g\,cm$^{-3}$] & 1.6 & +0.22 & -0.22 & isochrone fit \\
$\rho_{\star}$ [g\,cm$^{-3}$] & 2.7 & +1.05 & -1.05 & transit model \\
\hline
Planet : & & & & \\
\hline
$P$ [days]& 3.5787806 & +1.6$\times 10^{-7}$ & -1.6$\times 10^{-7}$ & \\
$T0$ [BJD] & 2454967.0304 & +4.4$\times 10^{-5}$ & -4.3$\times 10^{-5}$ & \\
$b$ & 0.291 & +0.05 & -0.05 & \\
$R_{\rm planet} / R_{\star}$ & 0.0997 & +0.00069 & -0.00075 & \\
$R$ [$R_{\rm Jup}$] & 0.96 & +0.05 & -0.05 & \\
$K$ [m\,s$^{-1}$] & 89.0 & +11.0 & -10.0 & \\
$e \cos \omega$ & -0.030 & +0.034 & -0.027 & \\
$e \sin \omega$ & 0.23 & +0.15 & -0.10 & \\
$M$ [$M_{\rm Jup}$] & 0.71 & +0.10 & -0.10 & \\
$a$ [AU] & 0.047 & +0.007 & -0.008 & \\
$\rho$ [g\,cm$^{-3}$] & 1.07 & +0.15 & -0.15 & \\
\hline
RV$_{\rm jitter}$ [m\,s$^{-1}$] & 17.6 & +8.0 & -6.1 & \\
\hline
\enddata
\end{deluxetable}

\begin{deluxetable}{lrrrr}
\tablecolumns{5}
\tablewidth{0pt}
\tablecaption{Parameters of the Kepler-43 (KOI-135) transiting system
\label{135t}}
\tablehead{
\colhead{Parameter [unit]} & {median} & {+1$\sigma$} & {-1$\sigma$} & {notes}
}
\startdata
$M_{\star}$ [$M_{\odot}$] & 1.23 & +0.04 & -0.04 & isochrone fit \\
$R_{\star}$ [$R_{\odot}$] & 1.34 & +0.055 & -0.055 & isochrone fit \\
log(g) & 4.27 & +0.029 & -0.029 & isochrone fit \\
$L_{\star}$ [$L_{\odot}$] & 2.12 & +0.22 & -0.22 & isochrone fit \\
$\rho_{\star}$ [g\,cm$^{-3}$] & 0.71 & +0.078 & -0.078 & isochrone fit \\
$\rho_{\star}$ [g\,cm$^{-3}$] & 0.74 & +0.067 & -0.067 & transit model \\
\hline
Planet : & & & & \\
\hline
$P$ [days] & 3.0240922  & +1.6$\times 10^{-7}$  & -1.6$\times 10^{-7}$ & \\
$T0$ [BJD] & 2454965.4169 & +6.7$\times 10^{-5}$ & -7.5$\times 10^{-5}$ & \\
$b$ & 0.631  & +0.025 & -0.027 &  \\
$R_{\rm planet} / R_{\star}$ & 0.0862  & +0.00048 & -0.00058 &  \\
$R$ [$R_{\rm Jup}$] & 1.12 & +0.047 & -0.047 & \\
$K$ [m\,s$^{-1}$] & 372.0  & +12.0 & -13.0 & \\
$e \cos \omega$ & -0.005 & +0.011 & -0.017 & \\
$e \sin \omega$ & 0.005 & +0.033 & -0.017 & \\
$M$ [$M_{\rm Jup}$] & 3.03  & +0.18 & -0.18 & \\
$a$ [AU] & 0.044 & +0.002 & -0.002 & \\
$\rho$ [g\,cm$^{-3}$] & 2.87 & +0.17 & -0.17 & \\
\hline
RV$_{\rm jitter}$ [m\,s$^{-1}$] & 30.0 & +11.0 & -8.0 & \\
\hline
\enddata
\end{deluxetable}

\begin{deluxetable}{lrrrr}
\tablecolumns{5}
\tablewidth{0pt}
\tablecaption{Parameters of the Kepler-423 (KOI-183) transiting system
\label{183t}}
\tablehead{
\colhead{Parameter [unit]} & {median} & {+1$\sigma$} & {-1$\sigma$} & {notes}
}
\startdata
$M_{\star}$ [$M_{\odot}$] & 1.07  & +0.05 & -0.05 & isochrone fit\\
$R_{\star}$ [$R_{\odot}$] & 0.99 & +0.054 & -0.054 & isochrone fit\\
log(g) & 4.48  & +0.04 & -0.04 & isochrone fit\\
$L_{\star}$ [$L_{\odot}$] & 0.96 & +0.16 & -0.16 & isochrone fit \\
$\rho_{\star}$ [g\,cm$^{-3}$] & 1.58 & +0.22 & -0.22 & isochrone fit \\
$\rho_{\star}$ [g\,cm$^{-3}$] & 1.65 & +0.36 & -0.21 & transit model \\
\hline
Planet : & & & & \\
\hline
$P$ [days] & 2.68432848 & +8.2$\times 10^{-8}$ & -8.2$\times 10^{-8}$ & \\
$T0$ [BJD] & 2454966.3548 & +2.6$\times 10^{-5}$ & -2.7$\times 10^{-5}$ & \\
$b$ & 0.11  & +0.10 & -0.08  & \\
$R_{\rm planet} / R_{\star}$ & 0.1242 & +0.00089 & -0.00037 &  \\
$R$ [$R_{\rm Jup}$] & 1.20 & +0.065 & -0.065 & \\
$K$ [m\,s$^{-1}$] & 101.0  & +13.0 & -14.0 & \\
$e \cos \omega$ & -0.002  & +0.039 & -0.048 & \\
$e \sin \omega$ & -0.010 & +0.043 & -0.068 & \\
$M$ [$M_{\rm Jup}$] & 0.72 & +0.12 & -0.12 & \\
$a$ [AU] & 0.0396 & +0.003 & -0.003 & \\
$\rho$ [g\,cm$^{-3}$] & 0.55 &  +0.09 & -0.09 & \\
\hline
RV$_{\rm jitter}$ [m\,s$^{-1}$] & 30.0 & +11.0 & -8.0 & \\
\hline
\enddata
\end{deluxetable}

\end{document}